\documentclass{modelica}
\hypersetup{
  pdftitle  = {Latex Template for the International Modelica Conference},
  pdfauthor = {Author Name1, Author Name2},
  pdfsubject = {14th International Modelica Conference 2021},
  pdfkeywords = {Modelica, conference, LaTeX, template},
  colorlinks,
  linkcolor=black,
  urlcolor=black,
  citecolor=black,
  pdfpagelayout = SinglePage,
  pdfcreator = pdflatex,
  pdfproducer = pdflatex}

\usepackage{tikz}
\usepackage{pgfplots}
\usepgfplotslibrary{fillbetween}
\usetikzlibrary{patterns}
\usetikzlibrary{datavisualization}
\usetikzlibrary{arrows}
\usepgfplotslibrary{groupplots}
\usepackage{paralist}
\usepackage{amssymb}

\begin{document}
\thispagestyle{empty}

\title{Analytical Treatment of Hollow Toroid Flux Tubes}
\author{Herbert Schmidt}
\affil{Department of Electrical Engineering, Bochum University of Applied Sciences, Germany {\small\texttt{herbert.schmidt@hs-bochum.de}}}

\maketitle\thispagestyle{empty} 
\abstract{
Stray flux tubes 
around cylindrical poles are commonly modelled starting from the results for planar flux tubes using the circumference of the cylinder as depth. While this is a tried and tested approach, we here discuss analytical expressions using the actual axisymmetric geometry of a fraction of a hollow torus and compare their results to those of the accepted approach.
}


\section{Introduction}
\begin{figure}[b]
\centering
\unitlength=0.01\columnwidth
\begin{picture}(100,49)
\put(-4,0){
\put(10,0){\includegraphics[width=0.8\columnwidth]{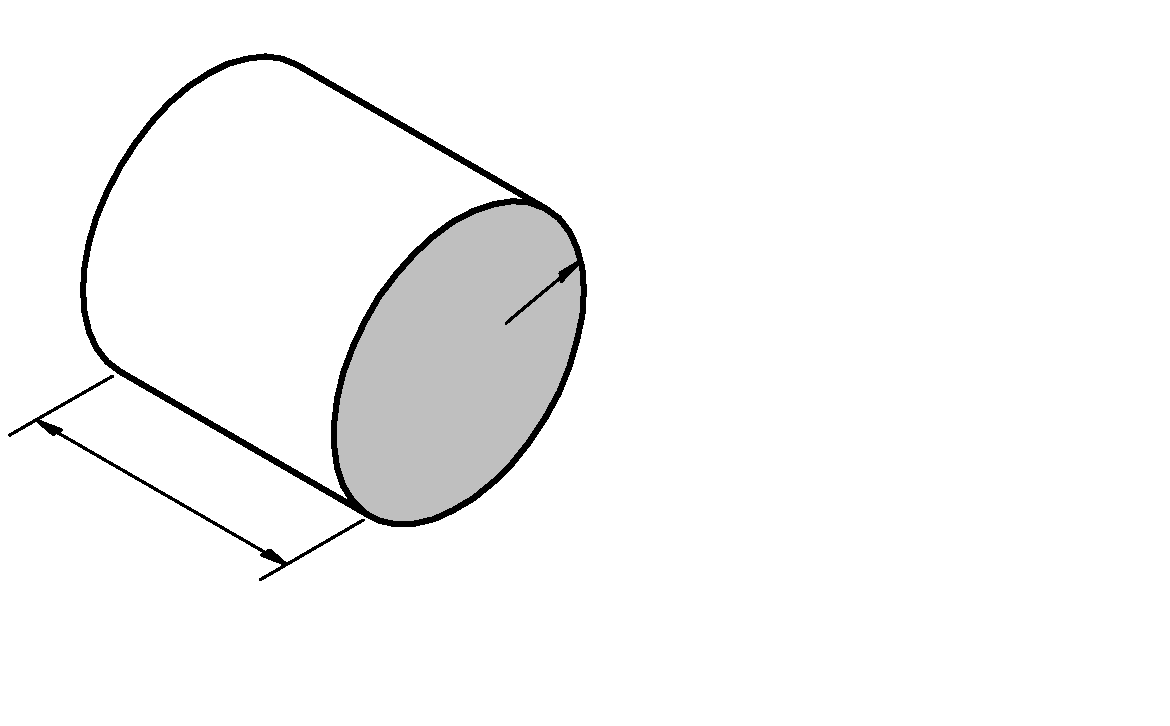}}
\put(18.5,13){$l$}
\put(43.5,30){$R$}
}
\put(1,0){
\put(10,0){\includegraphics[width=0.8\columnwidth]{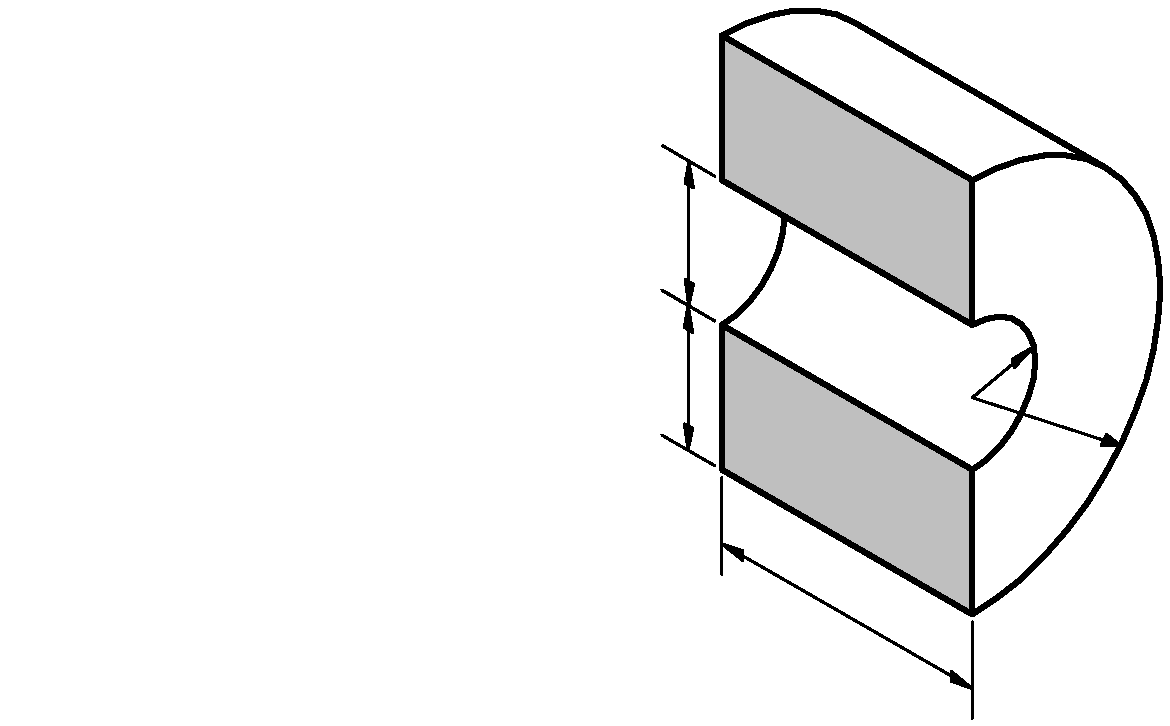}}
\put(54.2,33){$g$}
\put(54.5,23){$t$}
\put(75.2,24.5){$r_i$}
\put(82,21.5){$r_o$}
\put(65.8,4.2){$l$}
}
\put(0,0){(a)}
\put(50,0){(b)}
\end{picture}
\caption{Flux tube geometries. (a) generic prismatic flux tube with $A=\pi R^2$. (b) Half hollow cylinder with circumferential flux. In both figures, flux enters and exits the flux tube through areas marked in gray. }
\label{fig-1404-02}
\end{figure}
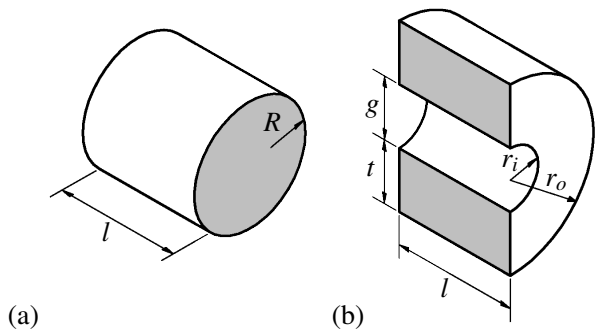

In analytically calculating the permeance $G_m$ (or equivalently the reluctance $R_m$) of a given stray flux tube, we start from Hopkinson's law for a prismatic element:
\begin{equation}
\frac{1}{R_m}=G_m=\frac{\Phi}{V_m}=\mu_0\frac{A}{l}
\end{equation}
where $\Phi$ is the flux perpendicular to a surface area $A$ and $V_m$ is the magnetic tension along its length $l$. This simple equation explicitely requires a prismatic flux tube, i.e., constant cross-section and constant length of flux lines within this element (cf.\ Fig.\ \ref{fig-1404-02}a). 

\begin{figure}
\centering
\unitlength=0.01\columnwidth
\begin{picture}(100,62)
\put(0,34){\includegraphics[scale=0.5]{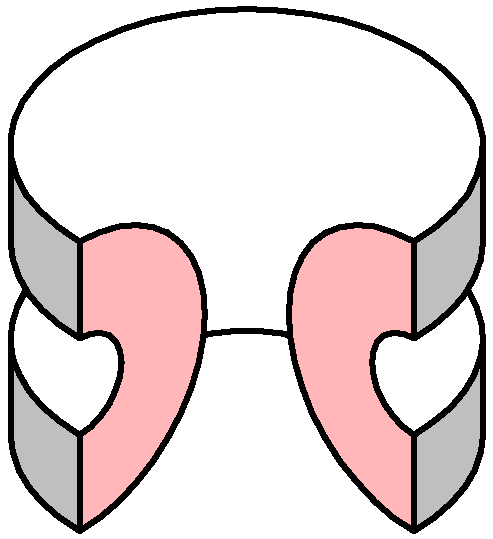}}
\put(22.5,17){\includegraphics[scale=0.5]{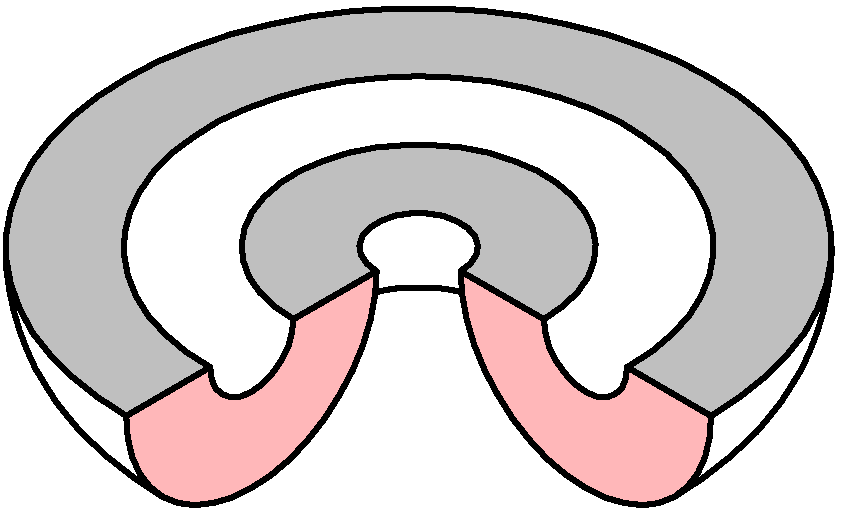}}
\put(58,32.8){\includegraphics[scale=0.5]{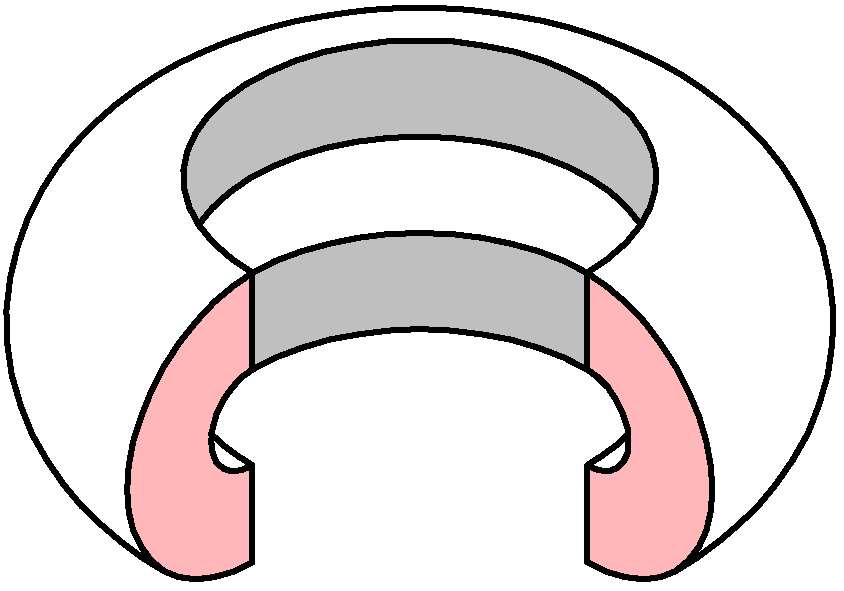}}
\put(0,1){\includegraphics[scale=0.5]{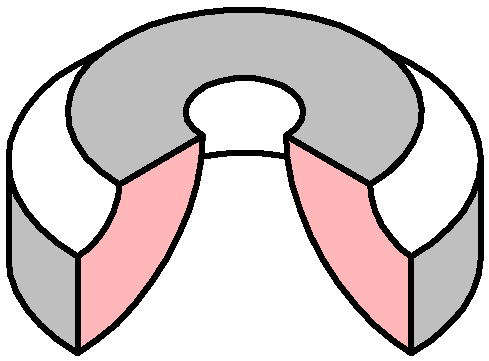}}
\put(58,0){\includegraphics[scale=0.5]{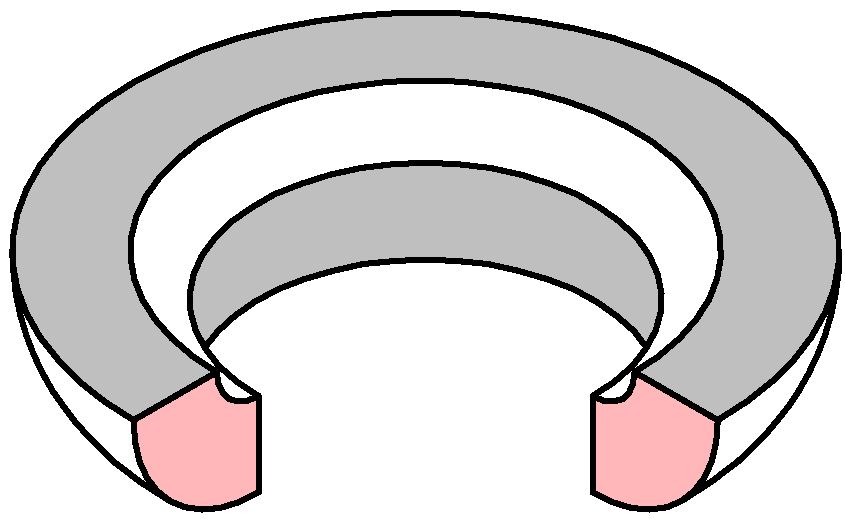}}
\put(9.5,34){(a)}
\put(41,17){(b)}
\put(77,34){(c)}
\put(9.5,0){(d)}
\put(77,0){(e)}
\end{picture}
\caption{Flux tube geometries. (a) inner half hollow torus. (b) lower half hollow torus. (c) outer half hollow torus. (d) inner quarter hollow torus. (e) outer quarter hollow torus. In all figures, flux enters and exits the flux tube through areas marked in gray. Red are areas opened by 3/4 cut-aways. The actual flux tubes extend over a full rotation.}
\label{fig-2304-01}
\end{figure}
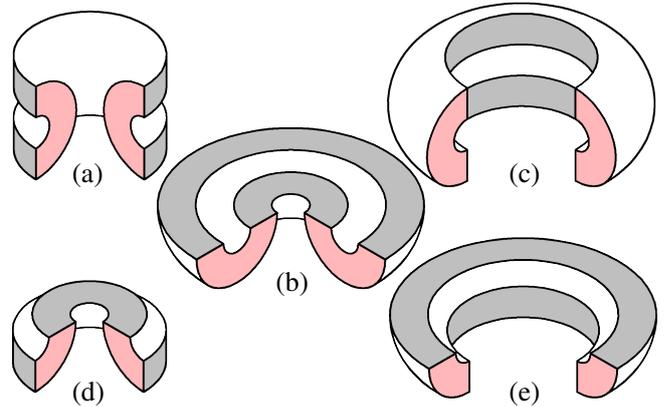

If one of these prerequisites is not met, the flux tube may be subdivided indefinitely and the effective permeance then is computed as an integral over these elements.\footnote{Note, that this is allowable only for stray flux tubes, i.e., $\mu_r=1$ (or at least constant), since variable cross section means variable operating point and hence variable permeability within the element.} This works well, e.g., for a hollow cylinder (or any fraction thereof) with radial or circumferential flux. A typical teaching problem might be a half hollow cylinder with circumferential flux (cf.\ Fig.\ \ref{fig-1404-02}b), resulting in \cite[P. 132, Eq. 8b]{Roters:1941}:
\begin{equation}
G_m=\frac{\mu_0 l}{\pi}\ln\frac{r_o}{r_i}=\frac{\mu_0l}{\pi}\ln\left(1+\frac{2t}{g}\right)
\label{eqn-1404-01}
\end{equation}
where $r_o$ is the outer radius, $r_i={g}/{2}$ is the inner radius, $t=r_o-r_i$ their difference, and $l$ is the length (depth) of the hollow cylinder. Here, the cross section $A=tl$ is constant, but the length of the flux lines varies from $\pi r_i$ to $\pi r_o$.

The same method that yields this analytical result becomes slightly more cumbersome, if both the cross section and the length vary within the flux tube. It still is valid, though, and in this paper we will show the analytical result for the flux in similar flux tubes wrapped within or without cylindrical poles, i.e., for half or quarter hollow torus geometries, where both the cross section per flux line and the length of each flux line are variable. We will cover five qualitatively different situations (cf.\ Fig.\ \ref{fig-2304-01}):
\vfill
\begin{compactitem}
\item[(a)] inner half hollow torus
\vfill
\item[(b)] lower (or upper) half hollow torus
\vfill
\item[(c)] outer half hollow torus
\vfill
\item[(d)] inner quarter hollow torus
\vfill
\item[(e)] outer quarter hollow torus
\end{compactitem} 
All these may be addressed by first focusing on the outer and inner half hollow torus. The quarter variants have half the reluctance of the half variants, while the lower (or upper) half hollow torus is the sum of an inner and an outer quarter hollow torus. 

Mind that these flux tubes look like half hollow cylinders in their cross section, but they are none, because they are wrapped around cylindrical poles. It is therefore not permissible to simply scale the result to arbitrary polar angles. Scaling from half to quarter hollow torus is the only acceptable fraction, since the flux tube is perfectly mirror symmetrical with respect to this cutting plane.

\section{Analytical Treatment}
\begin{figure}
\centering
\unitlength=0.01\columnwidth
\begin{picture}(100,85)
\put(15,0){\includegraphics[width=0.7\columnwidth]{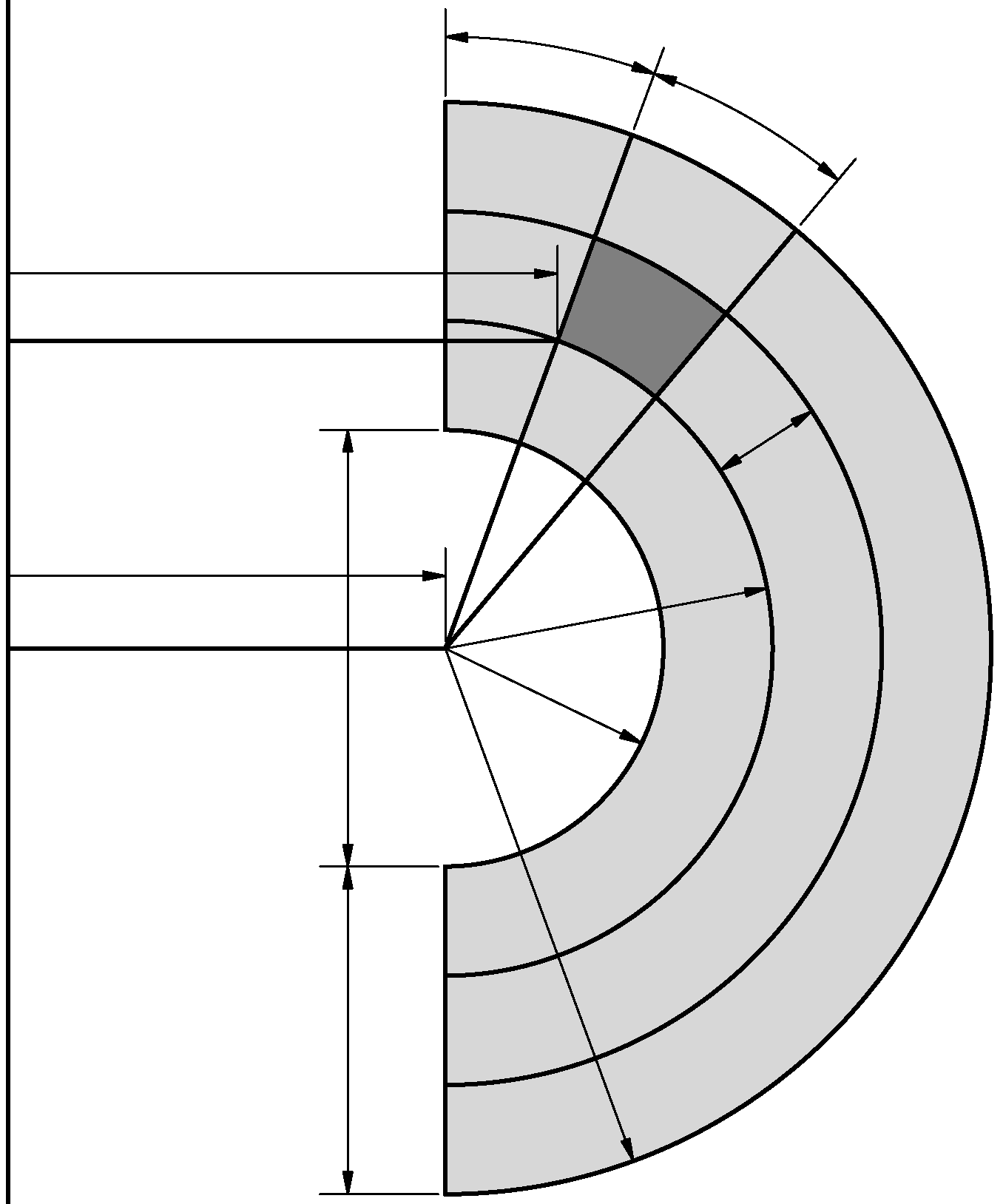}}
\put(29,44.8){$R$}
\put(22,66){$R+r\sin\vartheta$}
\put(53.5,20){$r_o$}
\put(53.5,36){$r_i$}
\put(36.5,35){$g$}
\put(37.2,11){$t$}
\put(52,82){$\vartheta$}
\put(66,77.5){${\rm d}\vartheta$}
\put(56,41.7){$r$}
\put(65.2,54){${\rm d}r$}
\end{picture}
\caption{Geometry used in integration.}
\label{fig-1504-01}
\end{figure}
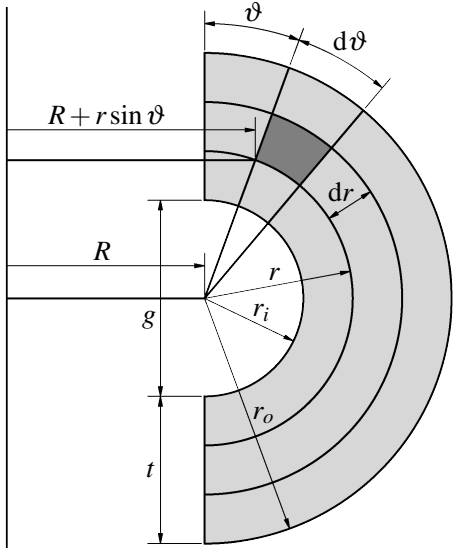

We use a subdivision as depicted in Figure \ref{fig-1504-01} and start by calculating the permeance of an infinitesimal flux tube at radius $r$ and polar angle $\vartheta$:
\begin{equation}
G_m(r,\vartheta)=\mu_0\frac{2\pi(R\pm r\sin\vartheta)\ {\rm d}r}{r\ {\rm d}\vartheta}
\end{equation}
The positive sign refers to the outer half hollow torus geometry shown in Figure \ref{fig-1504-01}, the negative sign refers to the corresponding inner half hollow torus.

We first derive the total permeance of a slice of polar width ${\rm d}\vartheta$:
\begin{align}
G_m(\vartheta)\ {\rm d}\vartheta&=\intop_{r_i}^{r_o}2\pi\mu_0\left(\frac{R}{r}\pm\sin\vartheta\right)\ {\rm d}r\\
&=2\pi\mu_0\left(R\ln\frac{r_o}{r_i}\pm(r_o-r_i)\sin\vartheta\right)
\end{align}
We then calculate the total reluctance of the full flux tube:
\begin{align}
R_m&=\frac{1}{2\pi\mu_0t}\intop_0^{\pi}\frac{{\rm d}\vartheta}{\frac{R}{t}\ln\frac{r_o}{r_i}\pm\sin\vartheta}
\end{align}
With $\eta=\frac{R}{t}\ln\frac{r_o}{r_i}$ we find two cases for the primitive \cite[P. 763, Eq. 306]{Bronstein:1993}:
\begin{equation}
\intop\frac{{\rm d}\vartheta}{\eta\pm \sin\vartheta}=\begin{cases}
\frac{2}{\sqrt{\eta^2-1}}\text{arctan}\frac{\eta\tan\frac{\vartheta}{2}\pm 1}{\sqrt{\eta^2-1}}&\text{for: } \eta>1\\
\frac{1}{\sqrt{1-\eta^2}}\ln\frac{\eta\tan\frac{\vartheta}{2}\pm 1-\sqrt{1-\eta^2}}{\eta\tan\frac{\vartheta}{2}\pm 1+\sqrt{1-\eta^2}}&\text{for: }\eta<1
\end{cases}
\end{equation}
We deviate from \textcite{Bronstein:1993} by stating the cases without the use of squares. Comparing $\eta$ to $1$ is equivalent to comparing $\eta^2$ to $1$ since all radii are positive and $r_o>r_i$, so $\eta$ is always positive.

\subsection{Inner Half Hollow Torus (Negative Sign)}
We first observe that only the case $\eta>1$ exists for the inner half hollow torus. To see this, we consider the case $\eta=1$:
\begin{align}
\frac{R}{t}\ln\frac{r_o}{r_i}&=1\\
\ln\frac{r_o/R}{r_i/R}&=\frac{r_o}{R}-\frac{r_i}{R}\\
\ln\frac{r_o}{R}-\frac{r_o}{R}&=\ln\frac{r_i}{R}-\frac{r_i}{R}\label{eqn-3004-001}
\end{align}
$\ln x-x$ is always negative, yet has a single high point at $x=1$. For the inner half hollow torus we know the argument to be smaller than one, otherwise the flux tube could not exist without intersecting itself. Thus this function is strictly monotonic increasing for relevant arguments, and the only solution of Equation \ref{eqn-3004-001} hence is $r_i=r_o$. In this case, the flux tube ceases to exist and $R_m\rightarrow\infty$. This means, the only relevant case is $\eta>1$.

Looking at the upper integration limit first, $\vartheta=\pi$. Then $\tan\frac{\vartheta}{2}\rightarrow+\infty$, while everything else in the argument of the arctan is finite, so the arctan returns $\frac{\pi}2$. Looking at the lower integration limit now, $\vartheta=0$. Then $\tan\frac{\vartheta}{2}=0$ and the argument of the arctan is negative, so we end up with:
\begin{align}
R_m
&=\frac{1}{\pi\mu_0t}\frac{1}{\sqrt{\eta^2-1}}\left(\frac{\pi}{2}+\text{arctan}\frac{1}{\sqrt{\eta^2-1}}\right)
\end{align}
using the point symmetry of the arctan. With $G_{m0}=\pi\mu_0t$ and using that the argument of the arctan is always positive, so we can use $\alpha_+=\frac{\pi}{2}+\text{arccot}\sqrt{\eta^2-1}$, we can write the permeance as:
\begin{align}
G_m&=G_{m0}\frac{\sqrt{\eta^2-1}}{\alpha_+}
\label{eqn-2804-001}
\end{align}
All major results, like this permeance, are collected in Table \ref{fig-2504-01} towards the end of the paper for quick reference.

We can finally verify the expected behaviour of the reluctance for $\eta^2\rightarrow 1$ by considering $\eta^2=1+\epsilon^2$:
\begin{equation}
\lim_{\epsilon\rightarrow0}\frac{1}{G_{m0}}\frac{1}{\epsilon}\left(\frac{\pi}{2}+\text{arctan}\frac{1}{\epsilon}\right)\rightarrow\infty
\end{equation}

\subsection{Outer Half Hollow Torus (Positive Sign)}
In turning to the outer half hollow torus now, we first note that in this case both $\eta>1$ and $\eta<1$ are physical. Consider first, $r_o=2r_i=R$, i.e., $t=r_i=R/2$ and $\eta=2\ln2>1$. Consider next $r_o=2r_i=2R$, i.e., $t=r_i=R$ and $\eta=\ln2<1$. The restriction for the inner half hollow torus exclusively came from the limited radial space towards the center -- outwards there is unlimited space.

Looking at the case $\eta>1$ first, the reasoning is much the same as before (here, $\alpha_-=\frac{\pi}{2}-\text{arccot}\sqrt{\eta^2-1}$, allowing for the sign), leaving:
\begin{align}
G_m&=G_{m0}\frac{\sqrt{\eta^2-1}}{\alpha_-}
\label{eqn-2804-002}
\end{align}
Looking at the case $\eta<1$ now, we find at $\vartheta=\pi$ that both the tangens in the numerator and the denominator of the logarithm's argument diverge, i.e. the argument approaches unity and the logarithm vanishes. At $\vartheta=0$, both tangens vanish, and we are left with:
\begin{equation}
R_m=-\frac{1}{2G_{m0}}\frac{1}{\sqrt{1-\eta^2}}\ln\frac{1-\sqrt{1-\eta^2}}{1+\sqrt{1-\eta^2}}
\label{eqn-3004-002}
\end{equation}
We use $\lambda=\ln\frac{1+\sqrt{1-\eta^2}}{1-\sqrt{1-\eta^2}}$ to write the permeance concisely:
\begin{equation}
G_m=G_{m0}\frac{2\sqrt{1-\eta^2}}{\lambda}
\label{eqn-2804-003}
\end{equation}
Note that the reciprocal in the logarithm's argument cancels the sign in Equation \ref{eqn-3004-002}. Both terms may not be evaluated at $\eta=1$, yet other than before they do not diverge but match up continuously. To see this, we start from the equation for $\eta>1$, and use $1=\eta^2-\epsilon^2$, arriving at:
\begin{equation}
\lim_{\epsilon\rightarrow0}\frac{1}{G_{m0}}\frac{1}{\epsilon}\left(\frac{\pi}{2}-\text{arccot}\,\epsilon\right)=\frac{1}{G_{m0}}
\end{equation}
Similarly, using $\eta^2=1-\epsilon^2$ for the equation for $\eta>1$ yields:
\begin{equation}
\lim_{\epsilon\rightarrow0}\frac{-1}{2G_{m0}}\frac{1}{\epsilon}\ln\frac{1-\epsilon}{1+\epsilon}=\frac{1}{G_{m0}}
\end{equation}
as above. So all these cases give physical solutions, and they match continuously.

\subsection{Inner and Outer Quarter Hollow Torus}
The equations for the inner and outer quarter hollow torus immediately follow from the observation, that their reluctance will be half that of the corresponding half hollow torus, their permeance will be twice that of the corresponding half hollow torus.
\subsection{Lower Half Hollow Torus}
The lower half hollow torus then follows from these:
\begin{align}
R_m&=\frac{\alpha_-}{2G_{m0}{\sqrt{\eta^2-1}}}+\frac{\alpha_+}{2G_{m0}{\sqrt{\eta^2-1}}}\\
G_m&=G_{m0}\frac{2\sqrt{\eta^2-1}}{\frac{\pi}{2}-\text{arccot}\sqrt{\eta^2-1}+\frac{\pi}{2}+\text{arccot}{\sqrt{\eta^2-1}}}\\
&=G_{m0}\frac{\sqrt{\eta^2-1}}{\pi/2}\qquad\text{for: }\eta>1
\end{align}
We do not need to consider other cases, as the inner quarter torus only exists for $\eta>1$.

\section{Check with Finite Element Method}
In order to evaluate the applicability of these formulae, we want to compare them with finite element method (FEM) calculations \cite{FEMM}. This is necessary since on our way here we made two separate steps: we firstly \textit{assumed} a flux pattern and we then secondly did the appropriate calculations to solve for this assumption. We did so far, however, not verify that the assumed flux patterns are correct in the first place. \textit{Are} the flux lines exactly shaped like half circles?
 The prior results are only valid insofar as this is a reasonable approximation of reality. 
 
 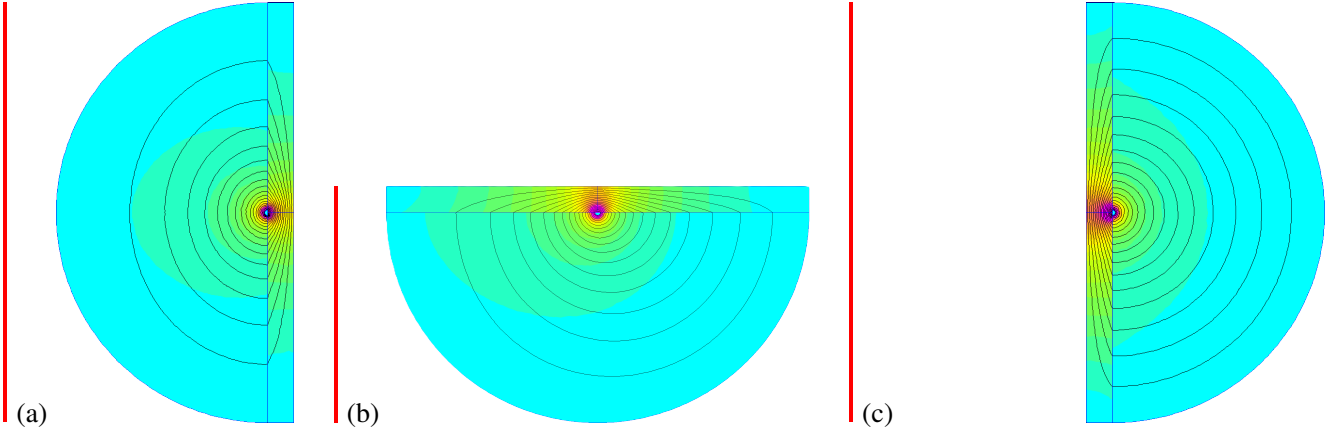
\begin{figure*} [t]
\centering
\unitlength=0.01\linewidth
\begin{picture}(100,32)
\put(0,0){\color{red}\rule{0.3\unitlength}{32\unitlength}}
\put(4,0){\includegraphics[height=32\unitlength]{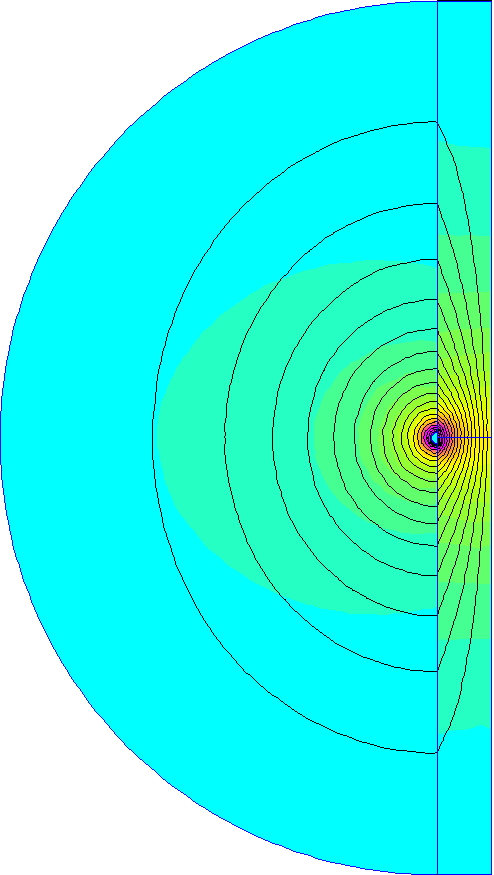}}
\put(25,0){\color{red}\rule{0.3\unitlength}{18\unitlength}}
\put(29,0){\includegraphics[width=32\unitlength]{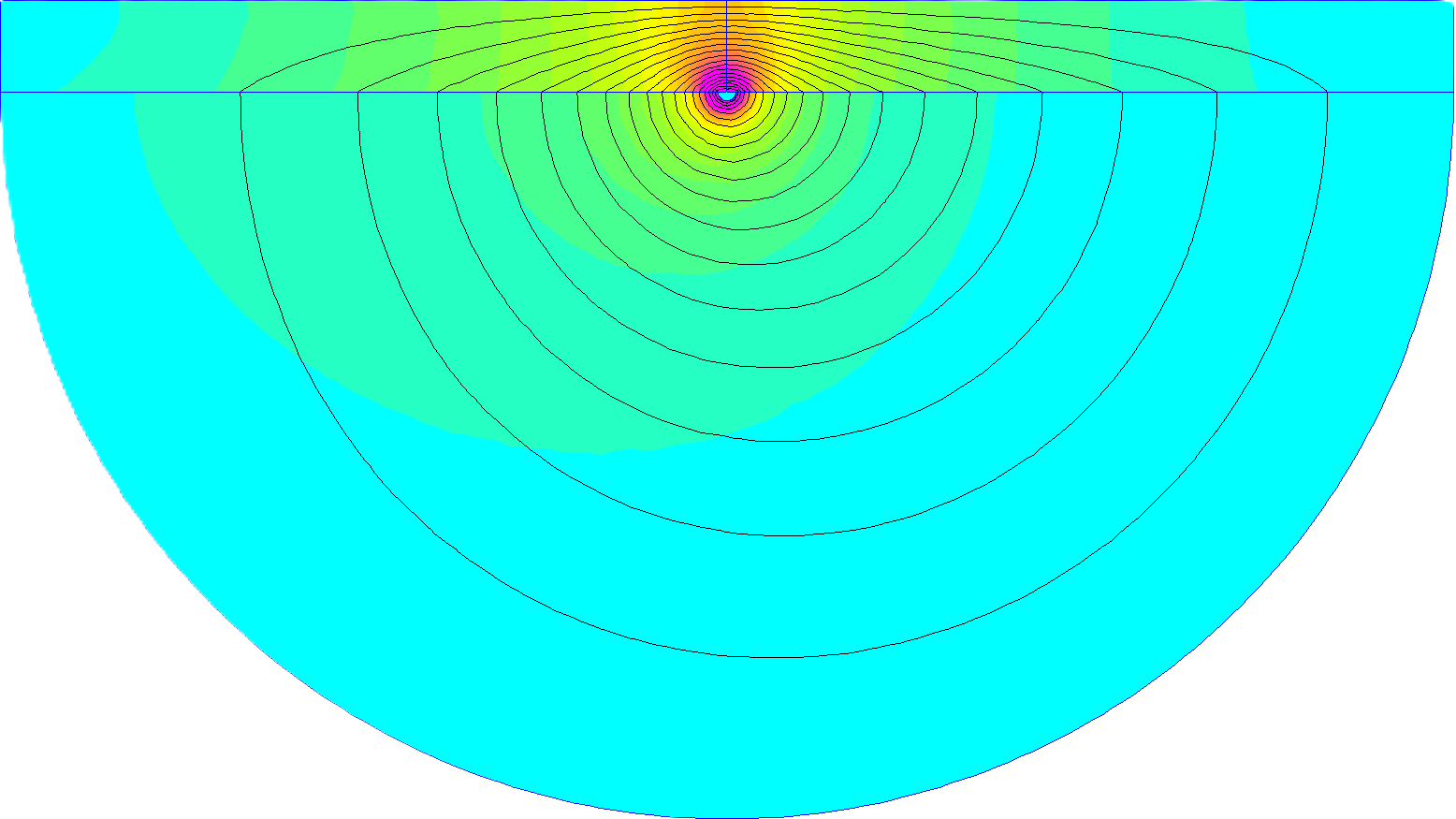}}
\put(64,0){\color{red}\rule{0.3\unitlength}{32\unitlength}}
\put(82,0){\includegraphics[height=32\unitlength]{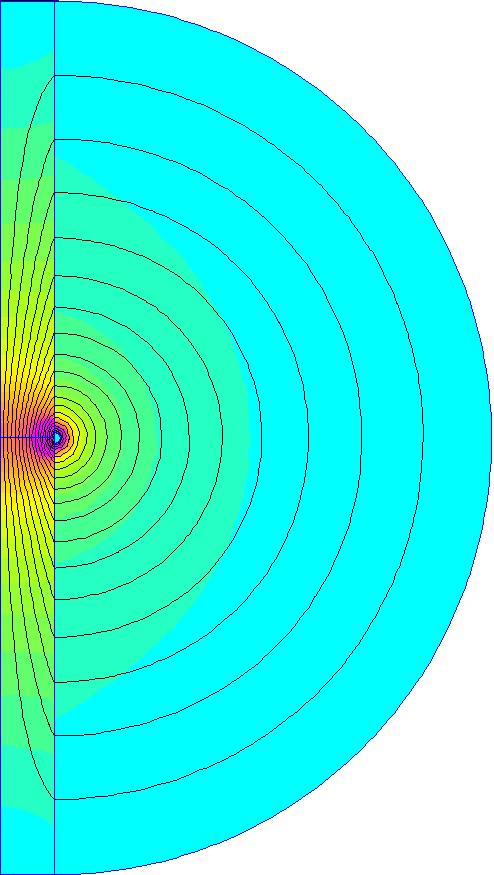}}
{
\put(1,0){(a)}
\put(26,0){(b)}
\put(65,0){(c)}
}
\end{picture}
\caption{FEM results for $r_o/R=0.8$ and $r_i/R=0.01$. (a) inner half hollow torus. (b) lower half hollow torus. (c) outer half hollow torus. The 
left vertical red lines indicate the rotational axes.}
\label{fig-2204-03}
\end{figure*}
 
To verify the flux pattern, we compare with FEM. The way the simulation is set up, only three areas are actually modelled (cf. Fig. \ref{fig-2204-03}): First, the flux tube in question, modelled as air ($\mu_r=1$). Second, the area inside $r_i$, modelled as an insulator ($\mu_r=10^{-6}$) containing a current linkage of, in this case, $\Theta=1$~A. Third, a yoke serving as a magnetic short ($\mu_r=10^{12}$). All outside areas are cut off by Dirichlet boundary conditions ($A=0$). We then evaluate the integral over $B_n$ in the center of the yoke, thus making sure that the flux lines at the point of integration have minimal curvature and are expected to yield a reliable value of $\Phi$. The permeance is finally computed from:
\begin{equation}
G_m=\frac{\Phi}{\Theta}
\end{equation}
Since all flux patterns scale, we effectively have to consider only two variables instead of three ($R$, $r_i$ and $r_o$). For FEM we fix $R=1$~mm for all calculations and use $r_i/R$ as independent variable for our plots. We show different values of $r_o/R$ as a family of curves, taking care to keep $r_i<r_o$ and $r_o<R$ where appropriate. Note that the quarter hollow torus elements (d) and (e) were not separately studied, as they are exactly half of (a) and (c). 

Three exemplary results of such analysis are shown in Figure \ref{fig-2204-03}. Apart from the permeance, these pictures give us valuable clues with respect to the previously raised question: \textit{are} the flux lines half-circle shaped? Even though we dictate the outer limit to be a half-circle, this is not exactly the case. In Figure \ref{fig-2204-03}a, e.g., it is quite evident that the distance of the outer-most flux line from the white surrounding area is \textit{not} constant. Similarly, it is obvious, that the false colour map (showing the absolute of the flux density) does not exactly mimic the circular symmetry of the flux tube. The teal area is horizontally elongated in Figure \ref{fig-2204-03}a, while it is vertically elongated in Figure \ref{fig-2204-03}c. So we do well to remain cautious.
 
\begin{figure*} [t]
\centering
\unitlength=0.01\linewidth
\begin{picture}(100,38)
\put(-2,0){
\put(9.2,0){(a)}
\put(40,0){(b)}
\put(70.8,0){(c)}
}
\put(0,0){\begin{tikzpicture}
    \begin{axis}[axis on top,
     width=0.38\textwidth,
     height=0.38\textwidth,
     xlabel={$r_i/R$},
     ylabel={$G_m/(\mu_0R)$ },
     xlabel near ticks,
     ylabel near ticks,
     every tick/.style={color=black, thin},
     xmode=log,
     ymode=log,
     minor x tick num=1,
     minor y tick num=1,
     xmin=0.01, xmax=4,
     ymin=0.0002, ymax=0.02,
    ]
   \addplot[color=black,mark=o,only marks] table[x expr=\thisrowno{0}, y expr=-\thisrowno{1}/(3.1415927*4e-7)] {out_i01.dat};
   \addplot[color=black,mark=o,only marks] table[x expr=\thisrowno{0}, y expr=-\thisrowno{1}/(3.1415927*4e-7)] {out_i02.dat};
   \addplot[color=black,mark=o,only marks] table[x expr=\thisrowno{0}, y expr=-\thisrowno{1}/(3.1415927*4e-7)] {out_i04.dat};
   \addplot[color=black,mark=o,only marks] table[x expr=\thisrowno{0}, y expr=-\thisrowno{1}/(3.1415927*4e-7)] {out_i08.dat};
   \addplot[color=blue,thick] table[x expr=\thisrowno{0}*1000, y expr=\thisrowno{1}] {eqn-ra01.txt};
   \addplot[color=blue,thick] table[x expr=\thisrowno{0}*1000, y expr=\thisrowno{1}] {eqn-ra02.txt};
   \addplot[color=blue,thick] table[x expr=\thisrowno{0}*1000, y expr=\thisrowno{1}] {eqn-ra04.txt};
   \addplot[color=blue,thick] table[x expr=\thisrowno{0}*1000, y expr=\thisrowno{1}] {eqn-ra08.txt};
   \addplot[color=red,thick] table[x expr=\thisrowno{0}*1000, y expr=\thisrowno{1}] {eqn-i01.txt};
   \addplot[color=red,thick] table[x expr=\thisrowno{0}*1000, y expr=\thisrowno{1}] {eqn-i02.txt};
   \addplot[color=red,thick] table[x expr=\thisrowno{0}*1000, y expr=\thisrowno{1}] {eqn-i04.txt};
   \addplot[color=red,thick] table[x expr=\thisrowno{0}*1000, y expr=\thisrowno{1}] {eqn-i08.txt};
     \end{axis}
\end{tikzpicture}}
\put(37.2,0){\begin{tikzpicture}
    \begin{axis}[axis on top,
     width=0.38\textwidth,
     height=0.38\textwidth,
     xlabel={$r_i/R$},
     xlabel near ticks,
     ylabel near ticks,
     every tick/.style={color=black, thin},
	yticklabel=\empty,
     xmode=log,
     ymode=log,
     minor x tick num=1,
     minor y tick num=1,
     xmin=0.01, xmax=4,
     ymin=0.0002, ymax=0.02,
    ]
   \addplot[color=black,mark=o,only marks] table[x expr=\thisrowno{0}, y expr=-\thisrowno{1}/(3.1415927*4e-7)] {out_u01.dat};
   \addplot[color=black,mark=o,only marks] table[x expr=\thisrowno{0}, y expr=-\thisrowno{1}/(3.1415927*4e-7)] {out_u02.dat};
   \addplot[color=black,mark=o,only marks] table[x expr=\thisrowno{0}, y expr=-\thisrowno{1}/(3.1415927*4e-7)] {out_u04.dat};
   \addplot[color=black,mark=o,only marks] table[x expr=\thisrowno{0}, y expr=-\thisrowno{1}/(3.1415927*4e-7)] {out_u08.dat};
   \addplot[color=blue,thick] table[x expr=\thisrowno{0}*1000, y expr=\thisrowno{1}] {eqn-ru01.txt};
   \addplot[color=blue,thick] table[x expr=\thisrowno{0}*1000, y expr=\thisrowno{1}] {eqn-ru02.txt};
   \addplot[color=blue,thick] table[x expr=\thisrowno{0}*1000, y expr=\thisrowno{1}] {eqn-ru04.txt};
   \addplot[color=blue,thick] table[x expr=\thisrowno{0}*1000, y expr=\thisrowno{1}] {eqn-ru08.txt};
   \addplot[color=red,thick] table[x expr=\thisrowno{0}*1000, y expr=\thisrowno{1}] {eqn-u01.txt};
   \addplot[color=red,thick] table[x expr=\thisrowno{0}*1000, y expr=\thisrowno{1}] {eqn-u02.txt};
   \addplot[color=red,thick] table[x expr=\thisrowno{0}*1000, y expr=\thisrowno{1}] {eqn-u04.txt};
   \addplot[color=red,thick] table[x expr=\thisrowno{0}*1000, y expr=\thisrowno{1}] {eqn-u08.txt};
     \end{axis}
\end{tikzpicture}}
\put(68,0){\begin{tikzpicture}
    \begin{axis}[axis on top,
     width=0.38\textwidth,
     height=0.38\textwidth,
     xlabel={$r_i/R$},
     xlabel near ticks,
     ylabel near ticks,
     every tick/.style={color=black, thin},
	yticklabel=\empty,
     xmode=log,
     ymode=log,
     minor x tick num=1,
     minor y tick num=1,
     xmin=0.01, xmax=4,
     ymin=0.0002, ymax=0.02,
    ]
   \addplot[color=black,mark=o,only marks] table[x expr=\thisrowno{0}, y expr=\thisrowno{1}/(3.1415927*4e-7)] {out_a01.dat};
   \addplot[color=black,mark=o,only marks] table[x expr=\thisrowno{0}, y expr=\thisrowno{1}/(3.1415927*4e-7)] {out_a02.dat};
   \addplot[color=black,mark=o,only marks] table[x expr=\thisrowno{0}, y expr=\thisrowno{1}/(3.1415927*4e-7)] {out_a04.dat};
   \addplot[color=black,mark=o,only marks] table[x expr=\thisrowno{0}, y expr=\thisrowno{1}/(3.1415927*4e-7)] {out_a08.dat};
   \addplot[color=black,mark=o,only marks] table[x expr=\thisrowno{0}, y expr=\thisrowno{1}/(3.1415927*4e-7)] {out_a16.dat};
   \addplot[color=black,mark=o,only marks] table[x expr=\thisrowno{0}, y expr=\thisrowno{1}/(3.1415927*4e-7)] {out_a32.dat};
   \addplot[color=blue,thick] table[x expr=\thisrowno{0}*1000, y expr=\thisrowno{1}] {eqn-ra01.txt};
   \addplot[color=blue,thick] table[x expr=\thisrowno{0}*1000, y expr=\thisrowno{1}] {eqn-ra02.txt};
   \addplot[color=blue,thick] table[x expr=\thisrowno{0}*1000, y expr=\thisrowno{1}] {eqn-ra04.txt};
   \addplot[color=blue,thick] table[x expr=\thisrowno{0}*1000, y expr=\thisrowno{1}] {eqn-ra08.txt};
   \addplot[color=blue,thick] table[x expr=\thisrowno{0}*1000, y expr=\thisrowno{1}] {eqn-ra16.txt};
   \addplot[color=blue,thick] table[x expr=\thisrowno{0}*1000, y expr=\thisrowno{1}] {eqn-ra32.txt};
   \addplot[color=red,thick] table[x expr=\thisrowno{0}*1000, y expr=\thisrowno{1}] {eqn-a01.txt};
   \addplot[color=red,thick] table[x expr=\thisrowno{0}*1000, y expr=\thisrowno{1}] {eqn-a02.txt};
   \addplot[color=red,thick] table[x expr=\thisrowno{0}*1000, y expr=\thisrowno{1}] {eqn-a04.txt};
   \addplot[color=red,thick] table[x expr=\thisrowno{0}*1000, y expr=\thisrowno{1}] {eqn-a08.txt};
   \addplot[color=red,thick] table[x expr=\thisrowno{0}*1000, y expr=\thisrowno{1}] {eqn-a16.txt};
   \addplot[color=red,thick] table[x expr=\thisrowno{0}*1000, y expr=\thisrowno{1}] {eqn-a32.txt};
     \end{axis}
\end{tikzpicture}}
\put(10.4,8){\includegraphics[scale=0.2]{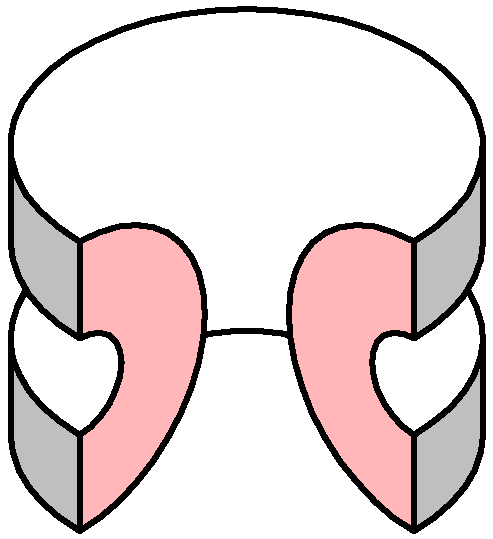}}
\put(41.2,8){\includegraphics[scale=0.2]{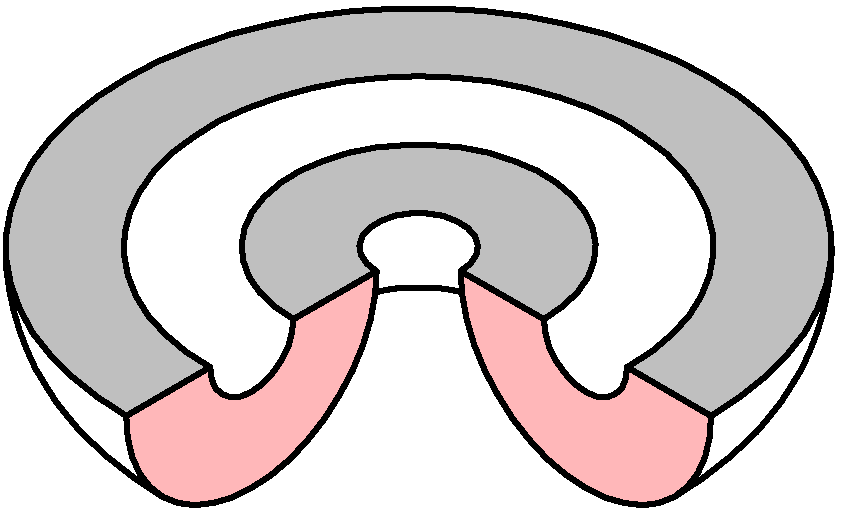}}
\put(72,8){\includegraphics[scale=0.2]{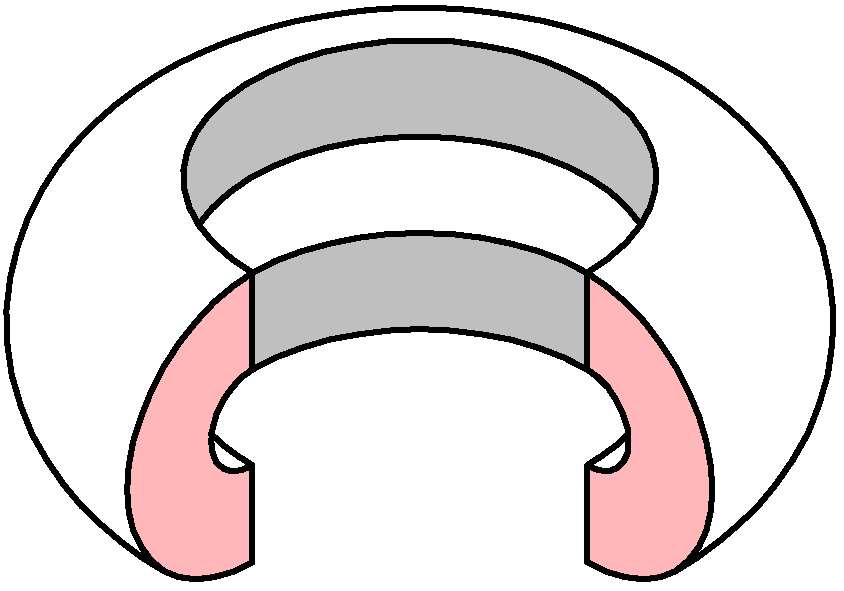}}
\end{picture}
\caption{Results for the permeance. (a) inner half hollow torus. (b) lower half hollow torus. (c) outer half hollow torus. Each left to right (bottom to top): $r_o/R=0.1$, 0.2, 0.4, 0.8, 1.6, 3.2 (the latter two only in (c)). Circles mark individual FEM results, red lines are analytical formulae presented here, blue lines are formulae presently used in Modelica.}
\label{fig-2204-01}
\begin{picture}(100,39)
\put(-2,0){
\put(9.2,0){(a)}
\put(40,0){(b)}
\put(70.8,0){(c)}
}
\put(0,0){\begin{tikzpicture}
    \begin{axis}[axis on top,
     width=0.38\textwidth,
     height=0.38\textwidth,
     xlabel={$r_i/R$},
     ylabel={absolute of relative deviation},
     xlabel near ticks,
     ylabel near ticks,
     every tick/.style={color=black, thin},
    xticklabel style={/pgf/number format/precision=2,/pgf/number format/fixed,/pgf/number format/fixed zerofill,
    /pgfplots/ticklabel shift=0.005\linewidth,/pgfplots/major tick length=0.008\linewidth,/pgfplots/minor tick length=0.004\linewidth},
     ymode=log,
	xticklabels={0.00,0.00,0.02,0.04,0.06,0.08},
     minor x tick num=1,
     minor y tick num=1,
     xmin=0, xmax=0.1,
     ymin=0.0001, ymax=0.1,
    ]
   \addplot[color=red,mark=*] table[x expr=\thisrowno{0}, y expr=\thisrowno{3}] {rel_vergl.txt};
   \addplot[color=blue,mark=o] table[x expr=\thisrowno{0}, y expr=\thisrowno{4}] {rel_vergl.txt};
     \end{axis}
\end{tikzpicture}}
\put(37.2,0){\begin{tikzpicture}
    \begin{axis}[axis on top,
     width=0.38\textwidth,
     height=0.38\textwidth,
     xlabel={$r_i/R$},
     xlabel near ticks,
     ylabel near ticks,
     every tick/.style={color=black, thin},
	yticklabel=\empty,
     xticklabel style={/pgf/number format/precision=2,/pgf/number format/fixed,/pgf/number format/fixed zerofill,
     /pgfplots/ticklabel shift=0.005\linewidth,/pgfplots/major tick length=0.008\linewidth,/pgfplots/minor tick length=0.004\linewidth},
     ymode=log,
	xticklabels={0.00,0.00,0.02,0.04,0.06,0.08},
     minor x tick num=1,
     minor y tick num=1,
     xmin=0, xmax=0.1,
     ymin=0.0001, ymax=0.1,
    ]
   \addplot[color=red,mark=*] table[x expr=\thisrowno{0}, y expr=\thisrowno{1}] {rel_vergl.txt};
   \addplot[color=blue,mark=o] table[x expr=\thisrowno{0}, y expr=\thisrowno{2}] {rel_vergl.txt};
     \end{axis}
\end{tikzpicture}}
\put(68,0){\begin{tikzpicture}
    \begin{axis}[axis on top,
     width=0.38\textwidth,
     height=0.38\textwidth,
     xlabel={$r_i/R$},
     xlabel near ticks,
     ylabel near ticks,
     every tick/.style={color=black, thin},
	yticklabel=\empty,
     xticklabel style={/pgf/number format/precision=2,/pgf/number format/fixed,/pgf/number format/fixed zerofill,
     /pgfplots/ticklabel shift=0.005\linewidth,/pgfplots/major tick length=0.008\linewidth,/pgfplots/minor tick length=0.004\linewidth},
     ymode=log,
	xticklabels={0.00,0.00,0.02,0.04,0.06,0.08},
     minor x tick num=1,
     minor y tick num=1,
     xmin=0, xmax=0.1,
     ymin=0.0001, ymax=0.1,
    ]
   \addplot[color=red,mark=*] table[x expr=\thisrowno{0}, y expr=\thisrowno{5}] {rel_vergl.txt};
   \addplot[color=blue,mark=o] table[x expr=\thisrowno{0}, y expr=\thisrowno{6}] {rel_vergl.txt};
     \end{axis}
\end{tikzpicture}}
\put(10.4,8){\includegraphics[scale=0.2]{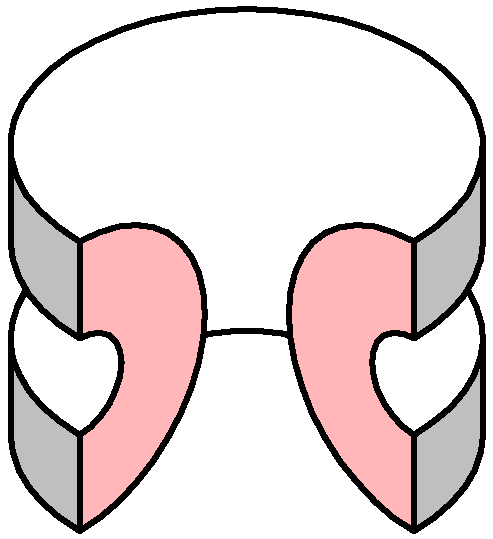}}
\put(41.2,8){\includegraphics[scale=0.2]{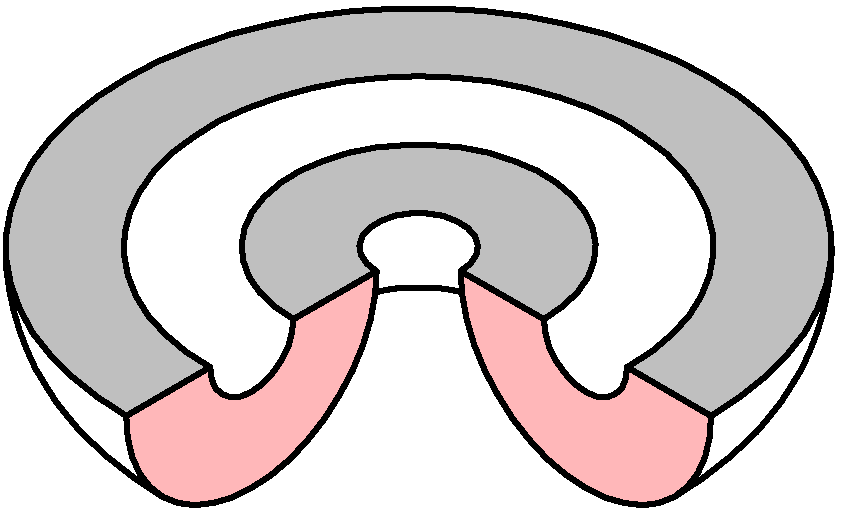}}
\put(72,8){\includegraphics[scale=0.2]{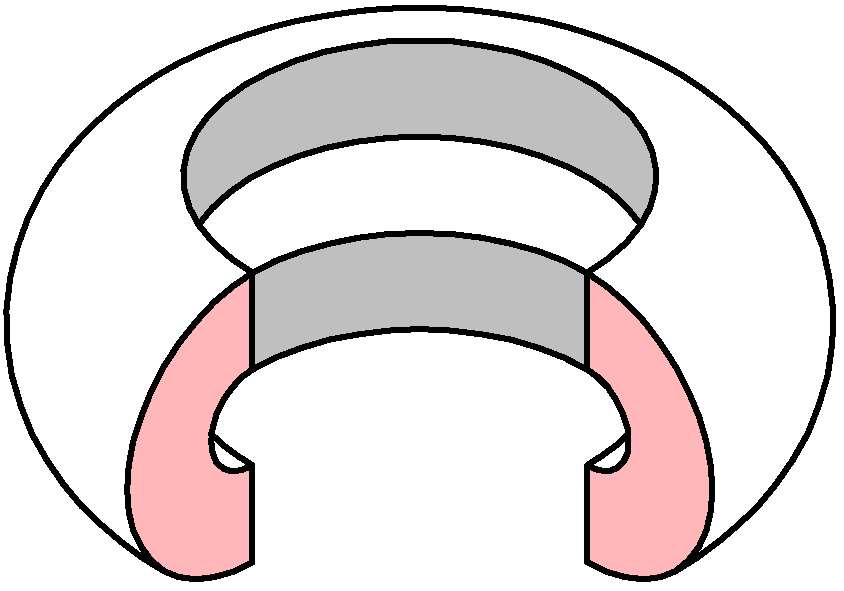}}
\end{picture}
\caption{Relative deviations from FEM results for $r_o/R=0.1$. (a) inner hollow hollow torus. (b) lower half hollow torus. (c) outer half hollow torus. Red filled symbols are analytical formulae presented here, blue open symbols are formulae presently used in Modelica.}
\label{fig-2204-02}
\end{figure*} 
  
Next we study the values for $G_m$ from FEM, from our formulae and from the formulae presently implemented in Modelica. The results are shown in Figure \ref{fig-2204-01}.

Let us consider the inner and outer half hollow torus first. Both these are compared to\linebreak \code{Modelica.\-Magnetic.\-Fluxtube.\-Shapes.Leakage. HalfHollowCylinder}, which essentially assumes that the half hollow torus is a half hollow cylinder wrapped around a cylinder:
\begin{equation}
G_m=2\mu_0R\ln\left(1+\frac{t}{r_i}\right)=G_{m0}\frac{\eta}{{\pi}/{2}}
\label{eqn-2204-01}
\end{equation}
The latter is written using the abbreviations introduced above. Comparing this to our expression for $G_{m}$ ($\eta>1$) requires $\sqrt{\eta^2-1}\rightarrow \eta$ in the numerator and $\text{arccot}\,\sqrt{\eta^2-1}\rightarrow 0$ in the denominator. Both are true for $\eta\rightarrow\infty$, i.e., the presently used Equation \ref{eqn-2204-01} holds up for:
\begin{equation}
\frac{R}{t}\ln\frac{r_o}{r_i}\rightarrow\infty
\end{equation}
In this limit, it similarly approaches our expressions for either half hollow torus (the sign of the arccot in the denominator is inconsequential, as in the considered limit the entire arccot is neglected). So our analytical result is not at odds with the present practice, but is a consistent extension to arbitrary parameter choices.

We intuitively expect Equation \ref{eqn-2204-01} to be fine for $R\gg r_i$ and $R\gg r_o$ (the former is noted in \code{Modelica.\-Magnetic.\-Fluxtube.\-Shapes.Leakage. HalfHollowCylinder} as a condition for cylindrical poles). What we find here is, that the actual condition is $R\gg(r_o-r_i)$ while at the same time $r_o$ markedly larger than $r_i$ (if the latter is not the case, the logarithm will be small). So the former condition cannot be fulfilled by making both $r_o$ and $r_i$ large while keeping their difference small. 

In Figure \ref{fig-2204-01}a and \ref{fig-2204-01}c we find a reasonable fit for $r_o/R=0.1$, however, for larger $r_o/R$ we find considerable deviations even for $r_i/R=0.01$. So if stating the above condition using a single simple equation, $R\gg r_o$ might be more useful than $R\gg r_i$ (since $r_i<r_o$, the former includes the latter anyway).

Considering the lower half hollow torus now, we find a quite reasonable fit with the equation presently used in Modelica \cite[P. 139, Eq. 22a]{Roters:1941},\footnote{There are two cases given in \cite{Roters:1941} which actually are used to decide which bit of open pole surface does not count towards the useful radius sector, depending on whether the inner or outer useful radius difference is larger. We do not note these cases here as we explicity only use the appropriate radius sectors here, by starting from $R$ and drawing two half circles with radii $r_i$ and $r_o$ centered there.} which after substituting the appropriate symbols actually is identical to Equation \ref{eqn-2204-01}.

This actually gives quite reasonable results with the exception of $r_o/R$ close to unity, since the systematic errors of the inner and outer bit mostly cancel each other out. Note however, that in this case, our exact result is not much more complex, yet gives consistently good results for all parameter combinations.

For all three cases, our analytical result shows excellent fit over the studied parameter range. There actually are deviations from exactly circular flux lines, however, these do not result in notable differences in permeance. In order to quantify this statement, we in Figure \ref{fig-2204-02} show the relative deviations at $r_o/R=0.1$. For high $r_o/R$ the shortcomings of the presently implemented approach are obvious. What we want to do here is to make sure that our analytical approach is at least as good as the presently used approach where there are no obvious problems with the latter. 

From this comparison we find that even for $r_o/R=0.1$ our approach is systematically better than the presently used formula. In the case of the inner and outer half hollow torus, it is better in terms of the relative deviation from the FEM results by at least one order of magnitude. In the case of the lower half hollow torus, the deviations partly cancel, especially for $r_i/R=0.1$ the result is almost as good as the one presented here. Still, we can conclude that over the studied parameter range the formulae presented here give consistently better results than the status quo.

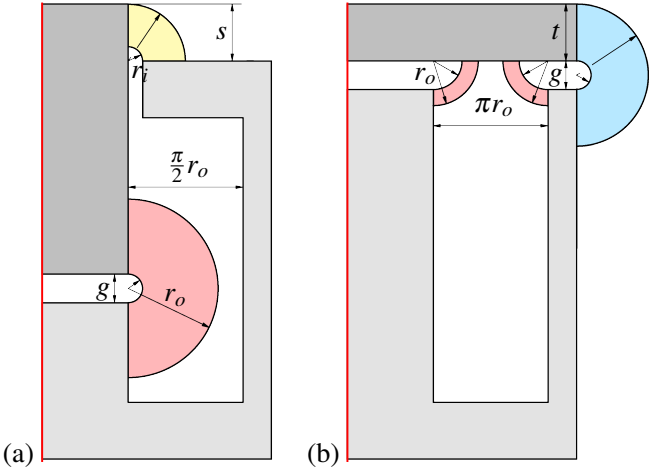
\begin{figure}
\centering
\unitlength=0.01\columnwidth
\begin{picture}(100,71.5)
\put(5.9,0){
\put(0,0){\includegraphics[width=1.0\columnwidth]{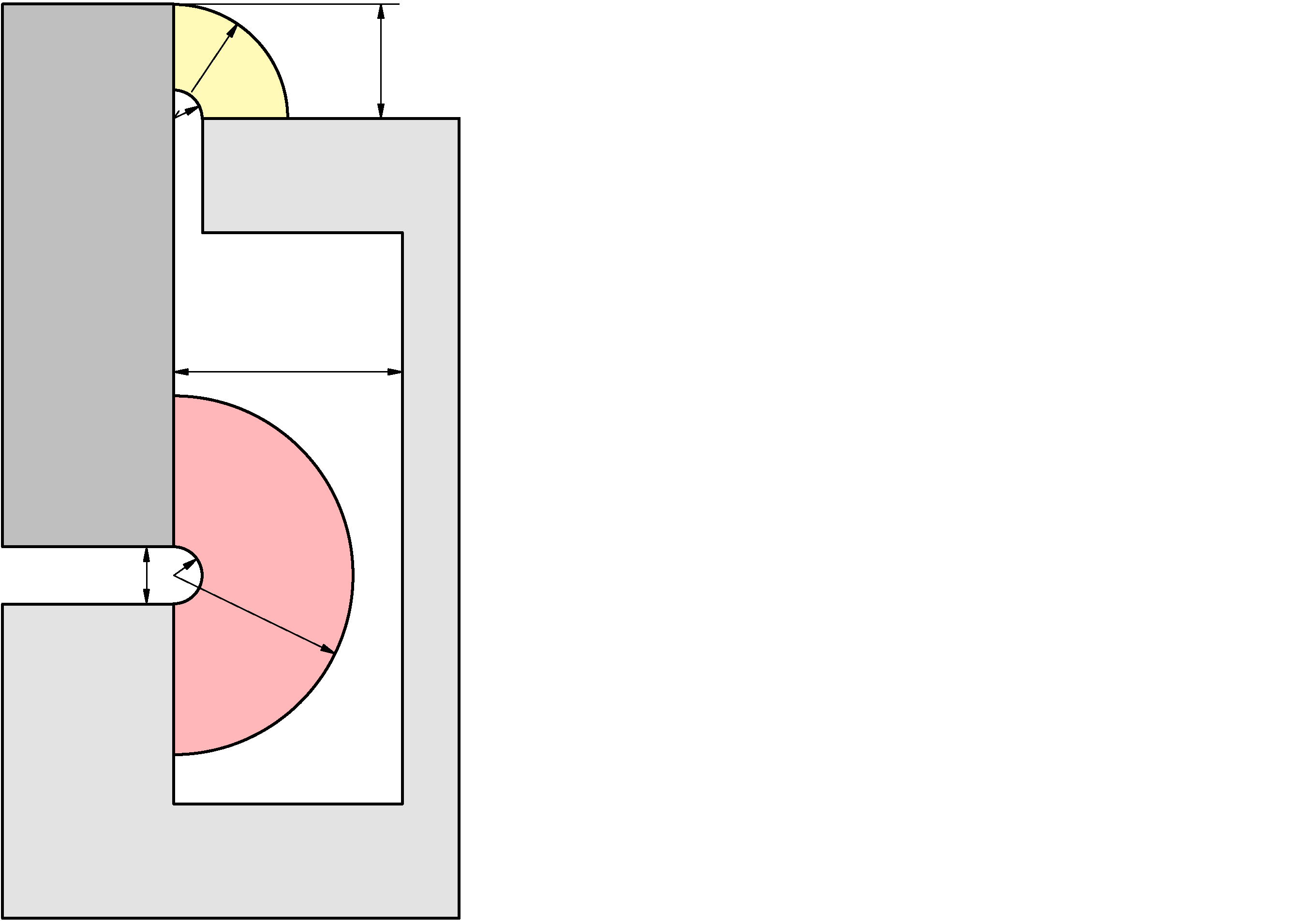}}
\put(0,0){\color{red}\rule{0.3\unitlength}{71.5\unitlength}}
\put(27,65.8){$s$}
\put(13.9,59.8){$r_i$}
\put(8.5,26){$g$}
\put(19.5,45){$\frac{\pi}{2}r_o$}
\put(19,25){$r_o$}
}
\put(0,0){(a)}
\put(0,0){\includegraphics[width=1.0\columnwidth]{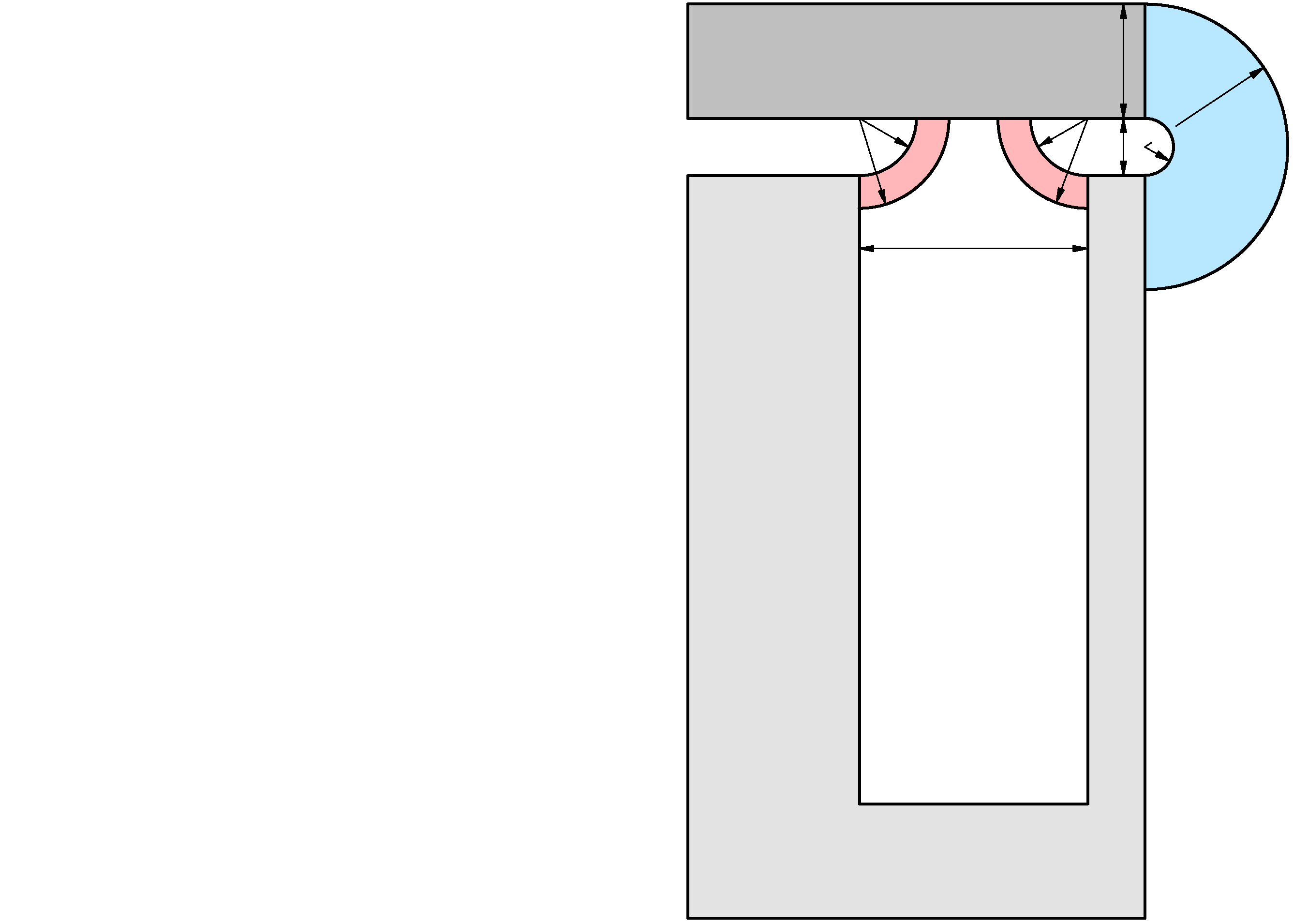}}
\put(53.05,0){\color{red}\rule{0.3\unitlength}{71.5\unitlength}}
\put(84.8,65.8){$t$}
\put(84.3,59.3){$g$}
\put(63.5,59.5){$r_o$}
\put(72.7,54.5){$\pi r_o$}
\put(47,0){(b)}
\end{picture}
\caption{Application cases for force calculation. (a) plunger type solenoid. (b) flat face type solenoid. Light gray is the yoke, dark gray the armature. The left vertical red lines indicate the rotational axes. Red flux tubes are of type $r_o=\text{const.}$, blue of type $t=\text{const.}$, yellow of type $r_i=\text{const.}$}
\label{fig-3004-01}
\end{figure}

\section{Equations for Force}
Having analytical expressions for the permeance of flux tube geometries (a) through (e) (cf.\ Fig.\ \ref{fig-2304-01}) puts us in a position to also provide analytical expressions  for forces generated by their deformation, e.g., by a moving armature. We use \cite[P. 197, Eq. 2a]{Roters:1941}:
\begin{equation}
F=\frac{1}{2}V_m^2\frac{{\rm d}G_m}{{\rm d}g}
\end{equation}
which in Modelica is implemented in \code{Modelica. Magnetic.\-FluxTube.\-BaseClasses.\-Force}. To utilize this, we only need to supply an equation for the latter derivative, \code{dGmBydx}. We will, however, have to address a number of different cases, as for computing the derivatives it is quintessential to specify, which quantities are allowed to change with the air gap, $g$. We will walk through the relevant cases to illustrate what is meant by this.

The first relevant case is $r_o=\text{const.}$ This case is appropriate if there is a natural limit to the radial extent of the flux tube. For an inner hollow torus this might simply be $r_o=R$. For an outer hollow torus this might be due to another part of the magnetic circuit, which would attract flux lines outwards from a given radius. A typical example might be the red flux tube in Figure \ref{fig-3004-01}a. At a point, where $\frac{\pi}{2}r_o$ equals the radius difference of the shown yoke parts, flux lines would short out to the outer part of the yoke rather than arcing over this outer half hollow torus. Analogous cases will occur for inner half hollow tori, if there are equivalent hollow cylinder parts facing each other. A similar example for an inner and outer quarter hollow torus might be seen in the red flux tubes in Figure \ref{fig-3004-01}b, where again the radius difference within the yoke gives an upper limit to what outer radius such a toroid flux tube could have.

The second relevant case is $t=\text{const.}$ This is the case presently implemented in 
\code{Modelica.\-Magnetic. FluxTubes.\-Shapes.\-Force.\-LeakageAroundPoles}. 
 This case is appropriate, if the stray flux tube is limited by a given axial width $t$ of either one or both of the facing elements. As an example, consider the blue flux tube in Figure \ref{fig-3004-01}b. While the armature moves, $r_i=g/2$ will change, $t$ will stay constant and $r_o=r_i+t$ will change accordingly. 
 
 Flux tubes of this type again may occur in the shape of inner or outer, half or quarter hollow tori.
  In discussing the relevant derivatives below, though, we will not consider the quarter hollow tori separately, as nothing qualitatively new happens when doing that.
 The force will simply be quadruple what we find in a half hollow torus, since for the quarter hollow torus twice the magnetic tension per length is applied (i.e., twice the magnetic field strength), while twice the distortion is effected on a quarter hollow torus for given stroke, $s$, as compared to a half hollow torus. Since for the derivatives, we only need the change of stroke, ${\rm d}s$, it is allowable to identify $r_i=s$ for $t=\text{const.}$ and $r_o=\text{const.}$, and $r_o=s$ for $r_i=\text{const.}$
  
The latter is the third relevant case, 
 and this case will mostly occur for inner or outer quarter hollow tori.\footnote{It is possible to invent scenarios, where this might occur for half hollow tori. These scenarios, however, appear to be rather artificial. For this reason, we do not present them here. If needed, the required formulae may easily be produced starting from what is presented here.}
 The technical situation might be a cylindrical plunger in a cylindrical hole, so an outer quarter hollow toroid flux tube is restricted in $r_o$ and $t$, while at the same time $r_i$ (being the fixed radius difference between plunger and hole) is constant. This situation is visualized by the yellow flux tube in Figure \ref{fig-3004-01}a. The same type of flux tube may occur on the inside, e.g., if  the face of the plunger is lowerd into a hole in the yoke.

Note, that since in axisymmetric geometries motion is expected along the rotational axis, lower half hollow toroid flux tubes will generally not generate force (while maintaining this general geometry). Looking at the case of two coaxial hollow cylinders moving relative to each other, e.g., the relevant flux tube would be expected to be significantly distorted away from the half hollow torus geometry considered here. We will therefore not give equations for these.

We will now proceed to look at the required derivatives. All the relevant results are collected in Table \ref{fig-2504-01} for quick reference (formulae for quarter tori are only shown for $r_i=\text{const.}$ to keep the table concise; the equations for $r_o=\text{const.}$ and $t=\text{const.}$ may easily be derived from those shown, as pointed out above).

\subsection{Inner Half Hollow Torus ($r_o=\text{const.}$)}
Starting from Equation \ref{eqn-2804-001}, the required derivative is of the structure:
\begin{equation}
\frac{{\rm d}G_m}{{\rm d}g}
=\left(\frac{G_m}{G_{m0}}\frac{{\rm d}G_{m0}}{{\rm d}r_i}+G_{m0}\frac{{\rm d}G_m/G_{m0}}{{\rm d}\sqrt{\eta^2-1}}\frac{{\rm d}\sqrt{\eta^2-1}}{{\rm d}\eta}\frac{{\rm d}\eta}{{\rm d}r_i}\right)\frac{{\rm d}r_i}{{\rm d}g}
\label{eqn-2404-001}
\end{equation}
wherein we need the following derivatives:
\begin{align}
\frac{{\rm d}G_{m0}}{{\rm d}r_i}&
=-\frac{G_{m0}}{t}\\
\frac{{\rm d}G_m/G_{m0}}{{\rm d}\sqrt{\eta^2-1}}&=\frac{\left(\frac{\pi}{2}+\text{arccot}\sqrt{\eta^2-1}\right)+\frac{\sqrt{\eta^2-1}}{\eta^2}}{\left(\frac{\pi}{2}+\text{arccot}\sqrt{\eta^2-1}\right)^2}\label{eqn-2404-003}\\
\frac{{\rm d}\sqrt{\eta^2-1}}{{\rm d}\eta}&=\frac{\eta}{\sqrt{\eta^2-1}}\\
\frac{{\rm d}\eta}{{\rm d}r_i}&
=\frac{1}{t}\left(\eta-\frac{R}{r_i}\right)\label{eqn-2404-002}\\
\frac{{\rm d}r_i}{{\rm d}g}&=\frac{1}{2}
\end{align}
This might look somewhat intimidating, however, there are a lot of recurring terms that are needed to calculate the permeance, anyway, so this can actually be coded quite efficiently. In doing that, it becomes necessary to ascertain $r_o>r_i$. Once this is no longer valid, this flux tube ceases to exist, resulting in $G_m=0$ and $\frac{{\rm d}G_m}{{\rm d}g}=0$. The same is true for the outer half hollow torus. This check is unnecessary if $t=\text{const.}$ is used, as this way $r_o-r_i=t$ for $t>0$ always results in an existing flux tube.

\subsection{Inner Half Hollow Torus ($t=\text{const.}$)}
The general procedure is much the same as above:
\begin{equation}
\frac{{\rm d}G_m}{{\rm d}g}
=G_{m0}\frac{{\rm d}G_m/G_{m0}}{{\rm d}\sqrt{\eta^2-1}}\frac{{\rm d}\sqrt{\eta^2-1}}{{\rm d}\eta}\frac{{\rm d}\eta}{{\rm d}r_i}\frac{{\rm d}r_i}{{\rm d}g}
\label{eqn-2404-004}
\end{equation}
The first term in Equation \ref{eqn-2404-001} may be omitted, as $G_{m0}=\pi\mu_0 t$ is constant in this case. $\eta$ we can therefore rewrite as:
\begin{equation}
\eta=\frac{R}{t}\ln\left(1+\frac{t}{r_i}\right)
\end{equation}
As far as the derivatives themselves are concerned, only Equation \ref{eqn-2404-002} needs to be adjusted accordingly:
\begin{equation}
\frac{{\rm d}\eta}{{\rm d}r_i}
=-\frac{R}{r_i(r_i+t)}
\end{equation}

\subsection{Outer Half Hollow Torus ($r_o=\text{const.}$)}
In this case, we need to keep in mind to separate the cases $\eta>1$, $\eta=1$ and $\eta<1$ (this was not necessary for the inner half hollow torus). 
Starting with $\eta>1$ and using Equation \ref{eqn-2804-002} this time, the equations look much the same as before for the inner half hollow torus and we can keep using Equation \ref{eqn-2404-001}. We only need to adjust Equation \ref{eqn-2404-003} to:
\begin{equation}
\frac{{\rm d}G_m/G_{m0}}{{\rm d}\sqrt{\eta^2-1}}=\frac{\left(\frac{\pi}{2}-\text{arccot}\sqrt{\eta^2-1}\right)-\frac{\sqrt{\eta^2-1}}{\eta^2}}{\left(\frac{\pi}{2}-\text{arccot}\sqrt{\eta^2-1}\right)^2}
\end{equation}
Since the permeance is defined using cases, we will have to check for continuity between these. In order to facilitate this, we will give the full relevant equation:
\small
\begin{align}
\frac{{\rm d}G_m}{{\rm d}g}
&=\left(\frac{G_m}{G_{m0}}\frac{{\rm d}G_{m0}}{{\rm d}r_i}+G_{m0}\frac{{\rm d}G_m/G_{m0}}{{\rm d}\sqrt{\eta^2-1}}\frac{{\rm d}\sqrt{\eta^2-1}}{{\rm d}\eta}\frac{{\rm d}\eta}{{\rm d}r_i}\right)\frac{{\rm d}r_i}{{\rm d}g}\\
&=\frac{G_{m0}}{2t\alpha_-}\left(\left(\frac{\eta}{\sqrt{\eta^2-1}}-\frac{1}{\eta^2\alpha_-}\right)\left(\eta-\frac{R}{r_i}\right)-\sqrt{\eta^2-1}\right)
\end{align}
\normalsize
For $\eta<1$  
 the following derivatives need to be added to our pool:
\begin{align}
\frac{{\rm d}G_m/G_{m0}}{{\rm d}\sqrt{1-\eta^2}}&=-2\frac{\frac{2\sqrt{1-\eta^2}}{\eta^2}-\ln\frac{1+\sqrt{1-\eta^2}}{1-\sqrt{1-\eta^2}}}{\ln^2\frac{1+\sqrt{1-\eta^2}}{1-\sqrt{1-\eta^2}}}\\
\frac{{\rm d}\sqrt{1-\eta^2}}{{\rm d}\eta}&=-\frac{\eta}{\sqrt{1-\eta^2}}
\end{align}
Again, we give the full relevant equation:
\small
\begin{align}
\frac{{\rm d}G_m}{{\rm d}g}
&=\left(\frac{G_m}{G_{m0}}\frac{{\rm d}G_{m0}}{{\rm d}r_i}+G_{m0}\frac{{\rm d}G_m/G_{m0}}{{\rm d}\sqrt{1-\eta^2}}\frac{{\rm d}\sqrt{1-\eta^2}}{{\rm d}\eta}\frac{{\rm d}\eta}{{\rm d}r_i}\right)\frac{{\rm d}r_i}{{\rm d}g}\\
&=\frac{G_{m0}2}{2t\lambda}\left(\left(\frac{2}{\eta\lambda}-\frac{\eta}{\sqrt{1-\eta^2}}\right)\left(\eta-\frac{R}{r_i}\right)-\sqrt{1-\eta^2}\right)
\label{eqn-24-04-5}
\end{align}
\normalsize
We now turn to consider the limit $\eta\rightarrow 1$, $\eta>1$, by Taylor expansion using $\eta^2=1+\epsilon^2\Rightarrow \sqrt{\eta^2-1}=\epsilon$ or $\eta=1+\frac{\epsilon^2}{2}$. First, we consider:
\begin{equation}
\alpha=\frac{\pi}{2}-\text{arccot}\,\epsilon\approx\epsilon-\frac{\epsilon^3}{3}=\epsilon\left(1-\frac{\epsilon^2}{3}\right)
\end{equation}
Since the factor in front of the brackets is the same for $\eta>1$ and $\eta<1$, we only consider the term in the brackets and 
 find 
 for $\epsilon\rightarrow0$:
\begin{equation}
-\frac{1}{3}\left(1+2\frac{R}{r_i}\right)
\end{equation}
We now turn to consider the limit $\eta\rightarrow 1$, $\eta<1$, again by Taylor expansion using $\eta^2=1-\epsilon^2\Rightarrow \sqrt{1-\eta^2}=\epsilon$ or $\eta=1-\frac{\epsilon^2}{2}$. First we consider:
\begin{align}
\lambda=\ln\frac{1+\epsilon}{1-\epsilon}&=\ln(1+\epsilon)-\ln(1-\epsilon)\\
&\approx2\epsilon+2\frac{\epsilon^3}{3}=2\epsilon\left(1+\frac{\epsilon^2}{3}\right)
\end{align}
We now again only consider the term in the brackets of Equation \ref{eqn-24-04-5} and finally arrive at the same expression:
\begin{equation}
-\frac{1}{3}\left(1+2\frac{R}{r_i}\right)
\end{equation}
The limits therefore coincide at $\eta=1$ and we can note in this case:
\begin{equation}
\frac{{\rm d}G_m}{{\rm d}g}=-\frac{G_{m0}}{2t}\frac{1}{3}\left(1+2\frac{R}{r_i}\right)
\end{equation}
This flux tube (like all others) only exists, if $r_o>r_i$. Since in this case, we keep $r_o$ constant while changing $g=2r_i$, it becomes important to handle the case $r_o\le r_i$ separately, resulting in $G_m=0$ and $\frac{{\rm d}G_m}{{\rm d}g}=0$.

\subsection{Outer Half Hollow Torus ($t=\text{const.}$)}

For the general structure of the solution we again use Equation \ref{eqn-2404-004}. Starting with $\eta>1$ and using Equation \ref{eqn-2804-002} we actually did compute all relevant derivatives before.
 Since we need to consider the limit $\eta=1$ in a minute, we note the full expression of this term:
\begin{equation}
\frac{{\rm d}G_m}{{\rm d}g}
=-\frac{G_{m0}R}{r_ir_o}\frac{\eta}{2\sqrt{\eta^2-1}}\frac{\hskip-2.2mm\left(\frac{\pi}{2}-\text{arccot}\sqrt{\eta^2-1}\right)-\frac{\sqrt{\eta^2-1}}{\eta^2}}{\left(\frac{\pi}{2}-\text{arccot}\sqrt{\eta^2-1}\right)^2}
\end{equation}
For $\eta<1$ we start from \ref{eqn-2804-003}, do not need any additional derivatives either, and we again note the full expression:
\begin{equation}
\frac{{\rm d}G_m}{{\rm d}g}
=-\frac{G_{m0}R}{r_ir_o}\frac{\frac{2}{\eta}-\frac{\eta}{\sqrt{1-\eta^2}}\ln\frac{1+\sqrt{1-\eta^2}}{1-\sqrt{1-\eta^2}}}{\ln^2\frac{1+\sqrt{1-\eta^2}}{1-\sqrt{1-\eta^2}}}
\end{equation}
In both cases, the first term is identical. In order to check for continuity, we check the remaining terms. First, $\eta^2=1+\epsilon^2\Rightarrow\sqrt{\eta^2-1}=\epsilon$ or $\eta=1+\frac{\epsilon^2}{2}$ as above:
\begin{equation}
\lim_{\epsilon\rightarrow0}\frac{\left(1+\frac{\epsilon^2}{2}\right)}{2\epsilon}\frac{\frac{\pi}{2}-\text{arccot}\,\epsilon-\frac{\epsilon}{1+\epsilon^2}}{\left(\frac{\pi}{2}-\text{arccot}\,\epsilon\right)^2}
=\frac{1}{3}
\end{equation}
Now for $\eta^2=1-\epsilon^2\Rightarrow\sqrt{1-\eta^2}=\epsilon$ or $\eta=1-\frac{\epsilon^2}{2}$, again using the limit for $\lambda$ stated earlier
  we find:
\begin{equation}
\lim_{\epsilon\rightarrow0}\frac{\frac{2}{1-\frac{\epsilon^2}{2}}-\frac{1-\frac{\epsilon^2}{2}}{\epsilon}\ln\frac{1+\epsilon}{1-\epsilon}}{\ln^2\frac{1+\epsilon}{1-\epsilon}}
=\frac{1}{3}
\end{equation}
As expected, the curve is continuous and for $\eta=1$ we can note:
\begin{equation}
\frac{{\rm d}G_m}{{\rm d}g}=-\frac{G_{m0}R}{3r_ir_o}
\end{equation} 

\subsection{Inner Quarter Hollow Torus ($r_i=\text{const.}$)}
Since the gap, $g$, is not meaningful for quarter hollow tori, we instead use the stroke, $s$, as discussed above (cf.\ Fig.\ \ref{fig-3004-01}a).
 A constant offset will not change the result, and we may simply use $r_o=s$ and $t=s-r_i$. We need to consider:
\begin{equation}
\frac{{\rm d}G_m}{{\rm d}s}=\frac{G_m}{G_{m0}}\frac{{\rm d}G_{m0}}{{\rm d}s}+G_{m0}\frac{{\rm d}G_m/G_{m0}}{{\rm d}\sqrt{\eta^2-1}}\frac{{\rm d}\sqrt{\eta^2-1}}{{\rm d}\eta}\frac{{\rm d}\eta}{{\rm d}s}
\label{eqn-2804-004}
\end{equation}
with:
\begin{equation}
\eta=\frac{R}{s-r_i}\ln\frac{s}{r_i}
\end{equation}
Most of the derivatives we have noted before. We only need:
\begin{align}
\frac{{\rm d}G_{m0}}{{\rm d}s}&=\frac{G_{m0}}{t}\\
\frac{{\rm d}\eta}{{\rm d}s}&=\frac{1}{t}\left(\frac{R}{r_o}-\eta\right)
\end{align}
We thus arrive at:
\small
\begin{align}
\frac{{\rm d}G_m}{{\rm d}s}
=\frac{2G_{m0}}{t\alpha_+}\left(\sqrt{\eta^2-1}+\left(\frac{\eta}{\sqrt{\eta^2-1}}+\frac{1}{\eta\alpha_+}\right)\left(\frac{R}{r_o}-\eta\right)\right)
\end{align}
\normalsize
Note that this term is positive, other than the terms considered before. This is due to the fact that by moving the plunger into the hole, the permeance in this case increases, so this flux tube (by itself) actually acts to push the plunger out of the yoke (it of course is more than compensated by the radial flux contribution pulling the plunger into the yoke - the latter term is well-described already, though, and therefore not included here).

This term only exists for $r_o<R$. Since we use $r_o=s$ while changing $s$, it becomes important to handle the case $s>R$ separately (the permeance remains that of $r_o=R$, the force, however vanishes, since the permeance does not change anymore). Furthermore, this term only exists for $s>r_i$ (resulting in $G_m=0$ and $\frac{{\rm d}G_m}{{\rm d}s}=0$).

\subsection{Outer Quarter Hollow Torus ($r_i=\text{const.}$)}

In this case we again need to consider cases depending on $\eta$. We start with $\eta>1$, can keep using Equation \ref{eqn-2804-004} and do not need any new derivatives. The desired result is:
\small
\begin{align}
\frac{{\rm d}G_m}{{\rm d}s}&=\frac{G_m}{G_{m0}}\frac{{\rm d}G_{m0}}{{\rm d}s}+G_{m0}\frac{{\rm d}G_m/G_{m0}}{{\rm d}\sqrt{\eta^2-1}}\frac{{\rm d}\sqrt{\eta^2-1}}{{\rm d}\eta}\frac{{\rm d}\eta}{{\rm d}s}\\
&=\frac{2G_{m0}}{t\alpha_-}\left(\sqrt{\eta^2-1}+\left(\frac{\eta}{\sqrt{\eta^2-1}}-\frac{1}{\eta\alpha_-}\right)\left(\frac{R}{r_o}-\eta\right)\right)
\end{align}
\normalsize
For $\eta<1$, no new derivatives are needed for arriving at:
\small
\begin{align}
\frac{{\rm d}G_m}{{\rm d}s}&=\frac{G_m}{G_{m0}}\frac{{\rm d}G_{m0}}{{\rm d}s}+G_{m0}\frac{{\rm d}G_m/G_{m0}}{{\rm d}\sqrt{1-\eta^2}}\frac{{\rm d}\sqrt{1-\eta^2}}{{\rm d}\eta}\frac{{\rm d}\eta}{{\rm d}s}\\
&=\frac{2G_{m0}}{t}\frac{2}{\lambda}\left(\sqrt{1-\eta^2}+\left(\frac{2}{\eta\lambda}-\frac{\eta}{\sqrt{1-\eta^2}}\right)\left(\frac{R}{r_o}-\eta\right)\right)
\end{align}
\normalsize
As before we need to consider the case $\eta=1$ by testing limits for both cases and coincidingly find:
\begin{equation}
\frac{{\rm d}G_m}{{\rm d}s}=\frac{2G_{m0}}{t}\left(1+\frac{2}{3}\left(\frac{R}{r_o}-1\right)\right)
\end{equation}

\section{Numerical Implementation}
All of these formulae have been implemented in Modelica \cite{GitHub}. This process is mostly straightforward, only a few remarks might be in place. 

Firstly, some flux tubes allow only for certain parameter combinations, and while the variable geometry changes, they may essentially cease to exist. 
 This might be fixed by setting $G_m=0$ and $\frac{{\rm d}G_m}{{\rm d}x}=0$ in those cases, however, since in the base class \code{Force} a reluctance is produced by calculating the inverse of $G_m$, the latter needs to be set to an arbitrary small number rather than zero (we chose $G_m=10^{-15}$~H). It is in the nature of absolute values, that none will be small as compared to every other value, thus there is an inherent (if practically small) risk in doing that. 

Secondly, while mathematically the three given cases $\eta\lesseqqgtr1$ cover all eventualities, numerically one needs to keep some distance from $\eta=1$. 
 Otherwise, Modelica is needlessly forced to evaluate terms that analytically safely converge, yet numerically might introduce significant errors. Experimentally it is found that by reserving 
 $\eta\in[0.999999, 1.000001]$ to the solution strictly correct for $\eta=1$ only, such problems can be avoided. Since all three solutions merge 
  continuosly, no harm is done by this.

Thirdly, 
  what does the added accuracy and versatility cost in terms of computational time? 
    We put this to the test using our 
     \code{Outer\-Half\-Hollow\-Torus\-Constantt} and the presently implemented 
      \code{LeakageAroundPoles}. 

We will start by looking at the accuracy of the force calculation. While our class is exact within the assumption of circular flux lines, the presently implemented class assumes bending a straight quarter hollow cylinder into a quarter hollow torus. This restricts the meaningful parameter choices. An additional, practical problem with \code{LeakageAroundPoles} is, which circumference to use? The class calls for the mean circumference to use as width, $w$, however, even if one accepts the arithmetic mean to be applicable, this mean changes as $r_o=r_i+t$ increases with increasing gap, $g=2r_i$. Most likely, users will either put $2\pi R$, i.e., an obviously constant, yet systematically low value, or $2\pi(R+t/2)$, which might be expected to fit best for low values of the gap. Actually, the former gives quantitatively better results in our case, and we therefore went with $w=2\pi R$. Other parameters were $t=R=10$~mm.

\begin{figure}
\includegraphics[width=1.0\linewidth]{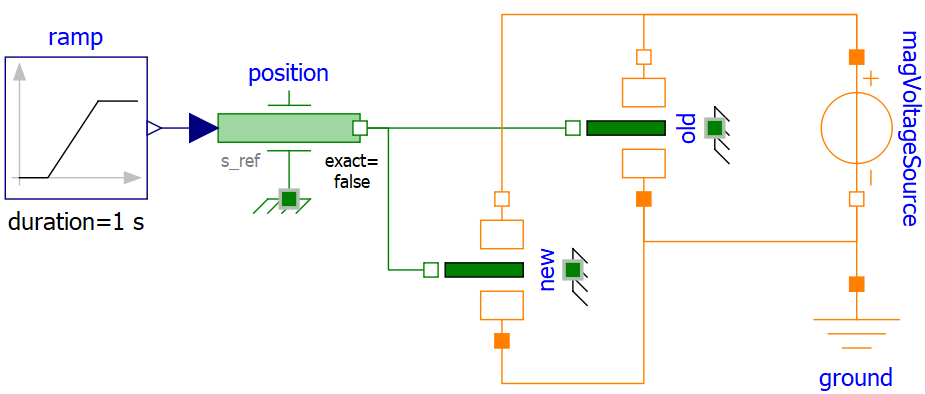}
\caption{OMEdit test model used to compare the ``new'' model to the established ``old'' model of an outer half hollow torus.}
\label{fig-1707025-01}
\end{figure}

\begin{figure}
\centering
\unitlength=0.01\linewidth
\begin{picture}(100,65.53)
\put(0,0){\begin{tikzpicture}
    \begin{axis}[axis on top,
     width=1.0\linewidth,
     height=0.713\linewidth,
     xlabel={$g$ in [mm]},     
     ylabel={relative deviation in [\%]},
     xlabel near ticks,
     ylabel near ticks,
     every tick/.style={color=black, thin},
     minor x tick num=4,
     minor y tick num=4,
     xmin=0, xmax=23,
     ymin=0, ymax=14,
    ]
   \addplot[color=black,thick] table[x expr=\thisrowno{0}*1000, y expr=100*(\thisrowno{2}-\thisrowno{1})/\thisrowno{1}] {old_vs_new_force_parallel.csv};
     \end{axis}
\end{tikzpicture}}
\end{picture}
\caption{Relative deviation of force calculated from ``old'' compared to ``new'' model. 
}
\label{fig-1707025-02}
\end{figure}
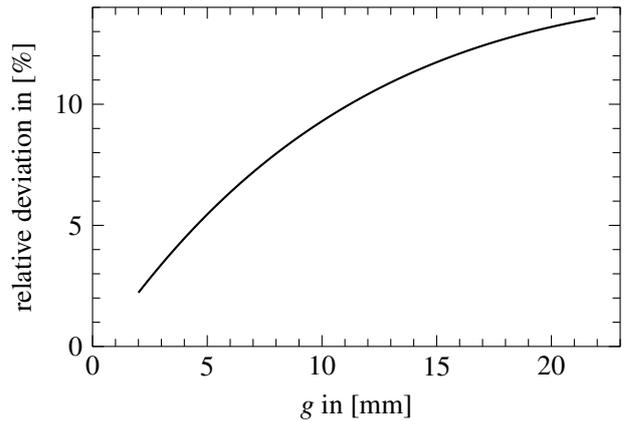

We set up a model (cf.\  Fig.\ \ref{fig-1707025-01}) using a prescribed positional ramp (20~mm stroke, 2~mm offset over 1~s) and a minimal magnetic loop including 1~A of magnetic tension to produce force vs.\ stroke curves. Figure \ref{fig-1707025-02} shows the relative deviation of these two curves. Note that the torus model went through all three cases $\eta\lesseqqgtr1$ within this ramp, without any trace of it in the result (as it should be). 
 
\code{LeakageAroundPoles} systematically neglects that the width of the flux tube is a function of the gap. Including 
 that would reflect into the derivative needed to calculate force and thereby change the class itself. This existing class deviates noticably from the more exact result. However, the significant relative deviation for large gaps is to be taken with a grain of salt, as absolute values of force in this region of stroke are very small. 

Finally looking at the computational times listed in the transformational debugger (in OMEdit), we find the following break down of major contributions:
\begin{compactitem}
\item 25.6~\% for the position
\item 13.3~\% for the old force calculation
\item 61.1~\% for the new force calculation
\end{compactitem}
At first glance, we find that our more elaborate model needs more than quadruple the computational time of the existing, simpler model. To put that into perspective: 12.2~\% of the computational time is used up in calculating $\ln\frac{r_o}{r_i}$ alone. 
 This illustrates that the absolute increase  in computational time is not quite as significant as it may appear from looking at the relative increase. 
  In absolute terms one may put it like this: the new force calculation is comparable in computational time to two position presets. 

Whether this is acceptable may obviously be decided on a case-to-case basis.
 In most geometries considered in this paper, there is no ``presently used model'' anyway, so mostly such flux tubes would have previously been neglected or falsly been modelled using the one existing class, even though that would not really have been applicable (e.g., if $r_o$ was constant rather than $t$).

\section{Conclusion}
In this paper we have derived exact expressions for the permeance  of half or quarter hollow toroid flux tubes (cf. Tab. \ref{fig-2504-01}). These expressions are an extension of the much simpler expression extensively used previously, lifting the limitation to a certain range of allowable parameters. Our derivation started from the assumption that the flux lines follow a general circular pattern. We verified our result by studying the actual flux patterns using FEM analysis. Quantitatively we find our expressions for the permeance to be in agreement with the FEM results within the expected limitations of the numerical method. We furthermore derived expressions necessary for calculating force, taking into account three relevant use cases (constant outer radius, constant radius difference, and constant inner radius of the torus, respectively). Corresponding Modelica models have been made available electronically.

\begin{table*}
\centering
\caption{Essential formulae in simulating hollow toroid flux tubes.}
\label{fig-2504-01}
\unitlength=0.01\linewidth
\begin{picture}(100,125)
\put(0,-25){
\put(0,12){
\put(0,0){\color{black}
\put(0,47){\line(0,1){91}}
\put(100,138){\line(-1,0){100}}
\put(100,138){\line(0,-1){91}}
}
\put(72,98){\line(0,1){40}}
\put(74,135.5){\sf\small\textbf{Symbols}}
\put(0,134.3){\line(1,0){100}}

\put(76.5,130){$\displaystyle
g=2r_i
$}

\put(76.9,124.4){$\displaystyle
t=r_o-r_i
$}

\put(74,118.8){$\displaystyle
G_{m0}=\pi\mu_0t
$}

\put(76.1,113.2){$\displaystyle
\eta=\frac{R}{t}\ln\frac{r_o}{r_i}
$}

\put(74.9,107.6){$\displaystyle
\alpha_\pm=\frac{\pi}{2}\pm\text{arccot}\,\sqrt{\eta^2-1}
$}

\put(76.1,102){$\displaystyle
\lambda=\ln\frac{1+\sqrt{1-\eta^2}}{1-\sqrt{1-\eta^2}}
$}

\put(2,135.5){\sf\small\textbf{Inner half hollow torus}}
\put(48.7,134.7){\includegraphics[scale=0.11]{Fig_a.png}}

\put(61,135.5){for $\eta>1$}
\put(18.1,130){$\displaystyle
G_m=G_{m0}\frac{\sqrt{\eta^2-1}}{\alpha_+}
$}

\put(2,122){for $r_o=\text{const.}$:}
\put(16.6,122){$\displaystyle
\frac{{\rm d}G_m}{{\rm d}g}=-\frac{G_{m0}}{2t\alpha_+}\left(\sqrt{\eta^2-1}-\left(\frac{\eta}{\sqrt{\eta^2-1}}+\frac{1}{\eta\alpha_+}\right)\left(\eta-\frac{R}{r_i}\right)\right)
$}

\put(2,114){for $t=\text{const.}$:}
\put(16.6,114){$\displaystyle
\frac{{\rm d}G_m}{{\rm d}g}=-\frac{G_{m_0}}{2r_o\alpha_+}\left(\frac{\eta}{\sqrt{\eta^2-1}}+\frac{1}{\eta\alpha_+}\right)\frac{R}{r_i}
$}

\put(0,110){\line(1,0){72}}
\put(2,107.5){\sf\small\textbf{Inner quarter hollow torus}}
\put(48.7,107.2){\includegraphics[scale=0.11]{Fig_d.png}}

\put(0,106.3){\line(1,0){72}}

\put(2,102){for $r_i=\text{const.}$:}
\put(61,107.5){for $\eta>1$}
\put(16.6,102){$\displaystyle
\frac{{\rm d}G_m}{{\rm d}s}
=\frac{2G_{m0}}{t\alpha_+}\left(\sqrt{\eta^2-1}+\left(\frac{\eta}{\sqrt{\eta^2-1}}+\frac{1}{\eta\alpha_+}\right)\left(\frac{R}{r_o}-\eta\right)\right)
$}

}

\put(0,110){\line(1,0){100}}
\put(2,107.5){\sf\small\textbf{Lower half hollow torus}}
\put(47.8,106.8){\includegraphics[scale=0.11]{Fig_b.png}}
\put(0,106.3){\line(1,0){100}}

\put(61,107.5){for $\eta>1$}
\put(18.1,102){$\displaystyle
G_m=G_{m0}\frac{\sqrt{\eta^2-1}}{\pi/2}
$}

\put(0,98){\line(1,0){100}}
\put(2,95.5){\sf\small\textbf{Outer half hollow torus}}
\put(47.8,94.6){\includegraphics[scale=0.11]{Fig_c.png}}
\put(0,94.3){\line(1,0){100}}

\put(18.1,86.5){$\displaystyle
G_m=\begin{cases}
G_{m0}\frac{\sqrt{\eta^2-1}}{\alpha_-} &\hskip34.5\unitlength\text{for: }\eta>1\\
G_{m0}{\color{white}\frac{\sqrt{\eta^2-1}}{\alpha_-}}							 &\hskip34.5\unitlength\text{for: }\eta=1\\
G_{m0}\frac{2\sqrt{1-\eta^2}}{\lambda} &\hskip34.5\unitlength\text{for: }\eta<1
\end{cases}
$}

\put(2,71){for $r_o=\text{const.}$:}
\put(16.6,71){$\displaystyle
\frac{{\rm d}G_m}{{\rm d}g}=\begin{cases}
-\frac{G_{m0}}{2t}\frac{1}{\alpha_-}\left(\sqrt{\eta^2-1}-
\left(\frac{\eta}{\sqrt{\eta^2-1}}-\frac{1}{\eta\alpha_-}\right)\left(\eta-\frac{R}{r_i}\right)\right)						&\hskip0.82\unitlength\text{for: }\eta>1\\
-\frac{G_{m0}}{2t}\frac{1}{3}\left(1+2\frac{R}{r_i}\right)						&\hskip0.82\unitlength\text{for: }\eta=1\\
-\frac{G_{m0}}{2t}\frac{2}{\lambda}\left(\sqrt{1-\eta^2}-\left(\frac{2}{\eta\lambda}-\frac{\eta}{\sqrt{1-\eta^2}}\right)\left(\eta-\frac{R}{r_i}\right)\right)						&\hskip0.82\unitlength\text{for: }\eta<1
\end{cases}
$}

\put(2,54.5){for $t=\text{const.}$:}
\put(16.6,54.5){$\displaystyle
\frac{{\rm d}G_m}{{\rm d}g}=\begin{cases}
-\frac{G_{m0}R}{r_ir_o}\frac{1}{2\alpha_-}\left(\frac{\eta}{\sqrt{\eta^2-1}}-\frac{1}{\eta\alpha_-}\right)						&\hskip20.3\unitlength\text{for: }\eta>1\\
-\frac{G_{m0}R}{r_ir_o}\frac{1}{3}					&\hskip20.\unitlength\text{for: }\eta=1\\
-\frac{G_{m0}R}{r_ir_o}\frac{1}{\lambda}\left(\frac{2}{\eta\lambda}-\frac{\eta}{\sqrt{1-\eta^2}}\right)					&\hskip20.3\unitlength\text{for: }\eta<1
\end{cases}
$}

\put(0,47){\line(1,0){100}}

\put(0,-51){
\put(2,95.5){\sf\small\textbf{Outer quarter hollow torus}}
\put(47.8,94.8){\includegraphics[scale=0.11]{Fig_e.png}}
\put(0,94.3){\line(1,0){100}}

\put(2,84.5){for $r_i=\text{const.}$:}
\put(16.6,84.5){$\displaystyle
\frac{{\rm d}G_m}{{\rm d}s}=\begin{cases}
\frac{2G_{m0}}{t}\frac{1}{\alpha_-}\left(\sqrt{\eta^2-1}+\left(\frac{\eta}{\sqrt{\eta^2-1}}-\frac{1}{\eta\alpha_-}\right)\left(\frac{R}{r_o}-\eta\right)\right)						&\hskip1.51\unitlength\text{for: }\eta>1\\
\frac{2G_{m0}}{t}\left(1+\frac{2}{3}\left(\frac{R}{r_o}-1\right)\right)				&\hskip1.51\unitlength\text{for: }\eta=1\\
\frac{2G_{m0}}{t}\frac{2}{\lambda}\left(\sqrt{1-\eta^2}+\left(\frac{2}{\eta\lambda}-\frac{\eta}{\sqrt{1-\eta^2}}\right)\left(\frac{R}{r_o}-\eta\right)\right)						&\hskip1.51\unitlength\text{for: }\eta<1
\end{cases}
$}

}
\put(0,25){\line(0,1){34}}
\put(100,25){\line(0,1){34}}
\put(0,25){\line(1,0){100}}

}
\end{picture}
\end{table*}

\section*{Acknowledgements}
The author would like to thank Silvia Hacia and Jörg Frochte for fruitful discussions, as well as Christof Kaufmann for technical support.

\printbibliography

\end{document}